\documentclass{article}

\usepackage{arXiv}

\usepackage[utf8]{inputenc} % allow utf-8 input
\usepackage[T1]{fontenc}    % use 8-bit T1 fonts

\usepackage{xcolor}
\usepackage[colorlinks = true,
            linkcolor = gray,
            urlcolor  = gray,
            citecolor = gray,
            anchorcolor = red]{hyperref}
            
\usepackage{url}            % simple URL typesetting

\usepackage{booktabs}       % professional-quality tables
\usepackage{multirow}       % multi-row tables
\usepackage{tabularx}       % advanced table formatting
\usepackage{makecell}       % better table cell formatting
\newcolumntype{C}{>{\centering\arraybackslash}X} % centered column type for tabularx

\usepackage{amsmath}        % advanced math typesetting
\usepackage{amsfonts}       % blackboard math symbols
\usepackage{nicefrac}       % compact symbols for 1/2, etc.

\usepackage{microtype}      % microtypography

\usepackage{apacite} % numbered citations
\usepackage{doi}            % DOI handling in bibliography

\usepackage{graphicx}       % graphics
\usepackage{float}          % control figure and table placement

\usepackage{fancyhdr}       % custom headers and footers
\title{AI-based Regional Emulation for Kilometer-Scale Dynamical Downscaling
    \author{
      \textbf{Yingkai Sha}$^{\dagger}$, %
      \textbf{Tracy Hertneky}$^{\dagger}$,
      \textbf{Ethan D. Gutmann}$^{\dagger}$,
      \textbf{Seth McGinnis}$^{\dagger}$,
      \textbf{Rachel McCrary}$^{\dagger}$,
      \\
      \textbf{Lulin Xue}$^{\dagger}$,
      \textbf{David John Gagne II}$^{\ddagger}$,
      \textbf{Kathryn Newman}$^{\dagger}$,
      \textbf{Andrew Newman}$^{\dagger}$\\[1em]
      Research Applications Laboratory\\
      NSF National Center for Atmospheric Research, Boulder, Colorado, USA$^{\dagger}$\\[1em]
      Computational and Information Systems Laboratory\\
      NSF National Center for Atmospheric Research, Boulder, Colorado, USA$^{\ddagger}$
    }
}
\begin{document}
\maketitle

\begin{abstract}
% <----------------- 250 word limit
An AI-based Limited-Area Model (LAM) is developed for dynamical downscaling over the Southern Great Plains and the southeastern United States, with strong generalization abilities under diverse boundary conditions. The model is trained using 0.25$^\circ$, 3-hourly ERA5 as forcings and CONUS404 as targets in 1980--2019, producing 4-km, hourly dynamical downscaling outputs; it is also connected to a post-processing model to derive additional diagnostic variables. The model is evaluated across multiple forcing datasets, time periods, and climate regimes. For present-day downscaling in water years 2021--2024, the model produces stable multi-year simulations with no unrealistic drift; its deterministic verification scores are comparable to other weather-forecasting-oriented AI models. The model also generalizes robustly to a 1.0$^\circ$, 6-hourly non-ERA5 forcing dataset, yielding only minor performance changes. Frontal cyclone and hurricane case studies further demonstrate that the model reconstructs realistic, interpretable weather-scale dynamical and thermodynamic structure from coarse boundary information. The AI-based LAM is further tested by downscaling 30-year global climate model runs in 1980--2010 and 2070--2100, and climate model ensembles in 2025-2027. In this application, the model remains stable at hourly downscaling frequencies for all 30 years and effectively captures climate-change signals, indicating meaningful generalization across different climate regimes. When downscaling ensembles, the model produces well-posed member distributions without collapsing the ensemble spread. Overall, the AI-based LAM of this study offers good downscaling performance and generalization abilities. It provides a practical and transferable example of adapting AI weather prediction models for regional climate applications.
\end{abstract}

%TC:ignore
\vspace{2ex}
\textbf{Plain Language Summary}
We built an artificial intelligence weather model to dynamically derive coarse atmospheric data into finer regional weather information over the Southern Great Plains and the southeastern United States. The model learns from a high-quality reference dataset and uses large-scale weather conditions as drivers to generate 4-km, hourly fields, plus additional variables from a post-processing step. We tested the system and found that it runs stably for multiple years without drifting to unrealistic values, and with good accuracy. The model works well when driven by a different, coarser dataset that it was not trained on. Case studies of a mid-latitude storm and a hurricane show that the model can reproduce key storm structures. The model was also applied to derive weather information from future climate simulations, including 30-year periods and ensemble runs; it captured climate-change signals well and preserved the predictive uncertainty from the original climate simulations.
%TC:endignore

% % 150 char limit, JGR submission only, will not appear in the arXiv
% \begin{keypoints}
% \item We introduce an AI-based limited-area model for 4-km, hourly dynamical downscaling over the Southern Great Plains and the southeastern US.
% \item The model is multi-year stable with good verification scores and works well across different boundary forcing datasets and temporal frequencies. 
% \item The model downscales climate model ensembles under historical and future regimes, capturing climate change signals while preserving ensemble spread.
% \end{keypoints}

\section{Introduction}

Accurate, high-resolution information on regional weather and climate plays a key role in the decision-making of water management, agriculture, and disaster risk reduction. Regional Climate Modeling (RCM) addresses this need through dynamical downscaling, which drives a Limited-Area Model (LAM) with boundary conditions from a coarser global dataset \cite<e.g.>{giorgi1991approaches,foley2010uncertainty,rummukainen2010state,feser2011regional,xue2014review,caldwell2009evaluation}. This strategy generates high-resolution, high-quality outputs, but is computationally expensive to apply at the temporal coverage and ensemble sizes required to project future climate conditions \cite{rampal2024enhancing,giorgi2015regional,prein2015review,kendon2017convection}.

In parallel, recent years have seen rapid progress in Artificial Intelligence (AI) Weather Prediction (AIWP). Global AIWP models have demonstrated competitive performance in medium-range forecasts comparable to that of Numerical Weather Prediction (NWP) models \cite<e.g.>{bi2023accurate,lam2023learning,chen2023fuxi,nguyen2023climax,bonev2023spherical,bodnar2024aurora,nguyen2023scaling,willard2024analyzing,lang2024aifs,schreck2025community}. By learning data-driven approximations of atmospheric dynamics from historical data, AIWP models can generate fast and skillful forecasts. Building on these advances, several studies have experimented with AI-based LAMs for regional weather forecasting. \citeA{nipen2025regional} proposed a 2.5 km LAM for 6-hourly forecasts up to 3 days. \citeA{abdi2025hrrrcast} introduced diffusion-model-based 6 km LAMs over CONUS, the HRRRCast system, for hourly ensemble forecasts up to 48 hours. \citeA{xu2025artificial} developed a 3 km LAM and ran it in parallel with Pangu Weather for hourly forecasts up to 48 hours. While these efforts primarily target short-range forecasts, their success suggests high potential for extending AI-based LAMs to dynamical downscaling, providing valuable high-resolution projections that have historically been computationally prohibitive to generate.

Dynamical downscaling and weather forecasting impose different requirements, and the role of AI-based LAMs on RCM remains under-explored. Compared to 1--10 day forecasts, dynamical downscaling requires stability over long-term integrations. AIWP models may exhibit minor distribution drifts \cite{jimenez2025ai,yuval2020stable}. This is acceptable in short-range forecasts, but can accumulate into unrealistic climatologies in a decades-long simulation. Dynamical downscaling also requires the AI-based LAM to generalize across different boundary forcings. In practice, these boundary conditions can come from reanalysis \cite<e.g.>{kotlarski2014regional,wang2020wrf,huang2018impact}, global NWP models \cite<e.g.>{lo2008assessment,shukla2013multi,sangelantoni2021dynamical}, or Earth system models \cite<e.g.>{wang2015high,bao2015dynamical}, with different resolutions, temporal frequencies, and climate regimes, all of which can influence the performance of AI-based LAMs. For model evaluation, many AIWP benchmarks emphasize short-range accuracy for instantaneous fields \cite{rasp2024weatherbench}, whereas RCM studies also care about long-term statistics such as mean state and spectral fidelity \cite<e.g.>{kotlarski2014regional,gutierrez2024performance}. These considerations indicate that developing new AI-based LAMs dedicated to dynamical downscaling is preferable to simply applying short-timescale AIWP models to a long-timescale problem. 

To bridge the gap in using AIWP models for RCM, this study develops and evaluates a novel AI-based LAM for dynamical downscaling. The proposed LAM is designed to produce multi-year, hourly, 4-km outputs by taking different boundary forcing resolutions (e.g., 0.25$^\circ$ and 1.0$^\circ$) and temporal frequencies (e.g., 3-hourly and 6-hourly) flexibly. The model is trained over a subset of the CONUS domain using  CONUS404 \cite{rasmussen2023conus404} data as the high-resolution target and the European Centre for Medium-Range Weather Forecasts (ECMWF) Reanalysis version 5 \cite<ERA5; >{hersbach2020era5} as baseline boundary forcings. Its training and evaluation prioritize long-term stability and generalization abilities. To demonstrate the effectiveness and practicality of the proposed AI-based LAM, we conduct and evaluate experiments across dynamical downscaling driven by multi-year reanalysis, global NWP outputs, and decadal climate prediction ensembles under both historical and future emission scenarios.   

This study addresses three scientific questions: (1) Can an AI-based LAM, motivated by state-of-the-art AIWP approaches, produce stable and skillful dynamical downscaling over multi-year periods? (2) How robustly can such a model generalize to boundary forcings that differ in resolution, frequency, and data source? (3) To what extent can a model trained under present-day climate generalize to future emission scenarios? By answering these questions, we aim to establish an AI-based LAM workflow for dynamical downscaling, quantify its strengths and limitations relative to RCM requirements, and provide guidance for future development of AIWP methods targeted at regional climate applications.

\section{Research domain and data}\label{sec2}

\subsection{Region of interest}\label{sec21}

\begin{figure}
    \centering
    \includegraphics[width=\columnwidth]{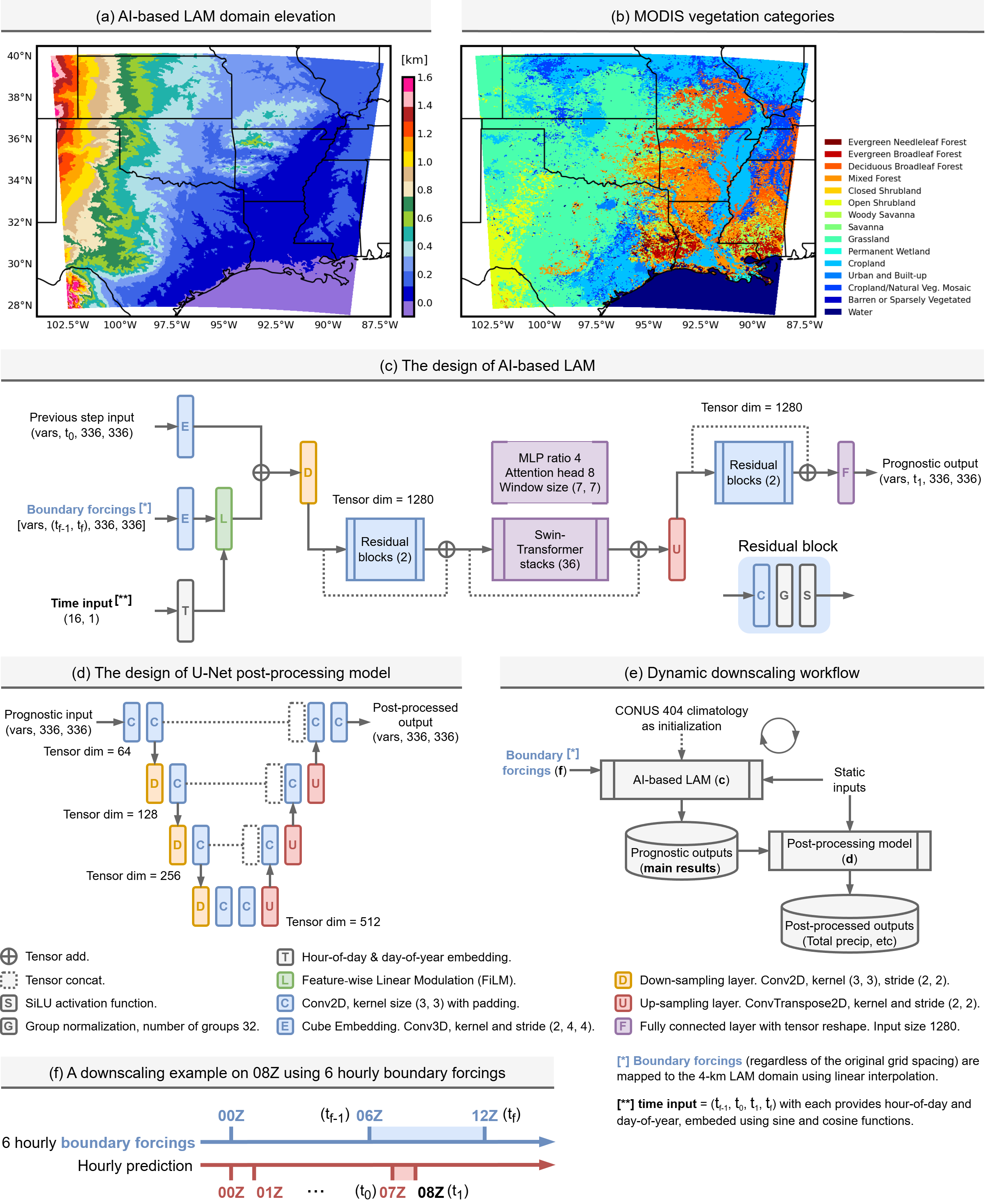}
    \caption{(a) The 4-km grid spacing domain with shaded elevation. (b) Vegetation categories of the domain. (c) Architecture of the AI-based LAM. (d) Architecture of the U-Net-based post-processing model. (e) The dynamical downscaling workflow. (f) An example of time window matching between hourly dynamical downscaling and 6-hourly boundary forcings.}
    \label{fig1}
\end{figure}

This study focuses on dynamical downscaling over the Southern Great Plains and the southeastern United States, spanning 28--40$^\circ$N and 87--103$^\circ$W (Figure~\ref{fig1}a). This area receives persistent low-level moisture transport from the Gulf. This includes nocturnal convection occurring over the Great Plains \cite{balling1985warm,reif201720,gebauer2018convection}, as well as sea breeze \cite{hill2010summertime,smith2005warm} and tropical cyclones influencing the Gulf Coast \cite{prat2013precipitation,maxwell2013tropical}, which bring intense precipitation and damaging winds. On inter-annual timescales, climate modes modulate the moisture availability of this area, contributing to swings between flooding and drought \cite{hu2001variations,mo2009influence}. In addition, this area contains large fractions of managed landscapes (cropland and grassland; Figure~\ref{fig1}b), linking regional climate variability to agricultural productivity and water-resource decision-making. Thus, dynamical downscaling over the Southern Great Plains and Southeastern United States is of high scientific value for demonstrating the ability of the AI-based LAM to resolve various smaller-scale systems, and of high priority for disaster risk reduction and climate-informed planning.

\subsection{Training Data}\label{sec22}

\begin{table}
\begin{center}
\caption{Summary of inputs and outputs in this study.}\label{tab1}
\renewcommand{\arraystretch}{1.2}
\begin{tabularx}{\textwidth}
{c c >{\centering\arraybackslash}X c c}
\specialrule{1.5pt}{0pt}{3pt}
Usage & Type & Variable Name & Units & Role \\ 
\midrule\vspace{-3ex}
\multirow{13}{*}{\makecell{AI-based LAM \\ 4-km input \\ and output}} & & & & \\ 
& \multirow{5}{*}{Upper air\textsuperscript{[a]}}
& Zonal wind       & $\mathrm{m \cdot s^{-1}}$   & \multirow{5}{*}{Instantaneous} \\
& & Meridional wind  & $\mathrm{m \cdot s^{-1}}$   & \\
& & Air temperature  & $\mathrm{K}$                & \\
& & Total water path & $\mathrm{kg \cdot kg^{-1}}$ & \\
& & Total pressure   & $\mathrm{Pa}$               & \\
\cmidrule(lr){2-5}
& \multirow{5}{*}{Single level}
& Surface pressure           & $\mathrm{Pa}$             & \multirow{5}{*}{Instantaneous} \\
& & 2-meter temperature      & $\mathrm{K}$               & \\
& & 10-meter zonal wind      & $\mathrm{m \cdot s^{-1}}$  & \\
& & 10-meter meridional wind & $\mathrm{m \cdot s^{-1}}$  & \\
& & Total precipitable water & $\mathrm{kg \cdot m^{-2}}$ & \\
\cmidrule(lr){2-5}
& \multirow{4}{*}{Others}
& Elevation             & $\mathrm{m}$   & \multirow{4}{*}{Static\textsuperscript{[c]}} \\
& & Orographic variance & $\mathrm{m^2}$ & \\
& & Vegetation category & -              & \\
& & Land-sea mask       & -              & \\
\midrule\vspace{-3ex}
\multirow{8}{*}{\makecell{AI-based LAM \\ boundary \\ forcing inputs}} & & & & \\ 
& \multirow{5}{*}{Upper air\textsuperscript{[b]}}
& Zonal wind          & $\mathrm{m \cdot s^{-1}}$    & \multirow{5}{*}{Instantaneous} \\
& & Meridional wind   & $\mathrm{m \cdot s^{-1}}$    & \\
& & Air temperature   & $\mathrm{K}$                 & \\
& & Specific humidity & $\mathrm{kg \cdot kg^{-1}}$  & \\
\cmidrule(lr){2-5}
& \multirow{4}{*}{Single level}
& Mean sea level pressure    & $\mathrm{Pa}$             & \multirow{4}{*}{Instantaneous} \\
& & 2-meter temperature      & $\mathrm{K}$              & \\
& & 10-meter zonal wind      & $\mathrm{m \cdot s^{-1}}$ & \\
& & 10-meter meridional wind & $\mathrm{m \cdot s^{-1}}$ & \\
\midrule\vspace{-3ex}
\multirow{5}{*}{\makecell{Post- \\ processed \\ outputs}} & & & & \\
& \multirow{2}{*}{Single level} 
& Soil moisture                & $\mathrm{m^3 \cdot s^{-3}}$ & \multirow{2}{*}{Instantaneous} \\
& & Soil temperature             & K                           & \\
\cmidrule(lr){2-5}
& \multirow{3}{*}{Flux form} 
& Total precipitation             & $\mathrm{mm}$              & \multirow{2}{*}{Cumulative} \\
& & Outgoing longwave radiation   & $\mathrm{W \cdot m^{-2}}$  & \\
\specialrule{1.5pt}{3pt}{0pt}
\end{tabularx}
\end{center}
\textsuperscript{a} Upper air variables for the AI-based LAM are on 12 hybrid sigma-pressure levels.\\
\textsuperscript{b} Upper air variables for the boundary forcings are on 11 constant pressure levels.\\
\textsuperscript{c} Static variables are input-only.\\
\end{table}

We train the AI-based LAM using CONUS404 as the high-resolution target and ERA5 as boundary forcings. CONUS404 provides an hourly, 4-km, 40-year analysis for North America, centered on CONUS. CONUS404 is dynamically downscaled from ERA5 using the Weather Research and Forecasting (WRF) model; it is verified to have good quality in resolving weather events and extremes \cite{rasmussen2023conus404,qin2025performance,wang2025hourly}. The \{0, 3, 6, 9, 12, 15, 18, 21, 24, 30, 36, 42\}-th WRF level of the CONUS404 over the region of interest is selected to construct the AI-based LAM; these levels span from roughly 1000 to 100 hPa. The ERA5 is selected on \{1000, 950, 850, 700, 600, 500, 400, 300, 200, 100, 50\} hPa constant pressure levels and interpolated linearly from 0.25$^\circ$ to the 4-km CONUS404 grid. ERA5 is chosen because it represents synoptic-scale circulations well, and it was the boundary forcing during the creation of CONUS404.

In addition to the autoregressive AI-based LAM, a post-processing step is proposed to derive additional variables. This model takes dynamically downscaled prognostic variables as inputs and produces post-processed variables such as total precipitation, on the same downscaling time step. This combination is inspired by HiRO-ACE \cite{perkins2025hiro}, in which a diffusion-model-based statistical downscaling model is one-way connected to a global AI climate model. The additional post-processing stage allows the AI-based LAM to focus on the dynamics of the core prognostic variables, thus providing flexibility to add more output variables based on user needs and creates space to extend the downscaling workflow with more sophisticated post-processing components in the future. The post-processing model is trained using the same CONUS404 dataset as the AI-based LAM, but with different variable sets. Table~\ref{tab1} summarizes the inputs and outputs for the AI-based LAM and the post-processing model.

For data pre-processing, CONUS404 and ERA5 are subset to 1980--2018 for training, and the year 2019 for validation. We downsample hourly ERA5 to 3-hourly. This allows the AI-based LAM to develop flexibility in handling different boundary forcing steps. We found that the AI-based LAM fails to handle longer boundary forcing steps if trained using paired hourly data, while if trained using 6-hourly or longer boundary steps, its downscaling performance is suboptimal. Using 3-hourly boundary forcing steps is a practical optimum that balances the generalization abilities and downscaling performance.

Variables are normalized using z-scores followed by residual normalization \cite{watt2023ace,sha2025investigating,schreck2025community}. Total water path and total column water are squared-rooted before z-scoring, and total precipitation is similarly quad-rooted. The means and standard deviations are computed from the training set (1980--2018). Exceptions to this normalization are the land-sea mask, which is converted to integer 0 and 1 values, and vegetation categories, which are scaled to $\mathrm{[0.0, 1.0]}$ (Figure~\ref{fig1}a and b).

\subsection{Testing Data}\label{sec23}
After training, we evaluate the performance of AI-based LAM using multiple boundary forcing datasets. ERA5 is used continuously as part of the test data at 3- and 6-hourly boundary forcing steps to assess the hourly dynamical downscaling performance of the model. The Global Data Assimilation System/Final Operational Global Analysis (GDAS/FNL) is selected as a testing dataset independent of ERA5. GDAS/FNL is provided at 6-hourly and 1.0$^\circ$ resolutions, representing coarser and operationally distinct boundary forcings for the dynamical downscaling of present-day conditions. The Community Earth System Model version 2 Large Ensemble (CESM-LENS2) is also applied for testing only. CESM-LENS2 is a set of ensemble climate simulations with historical runs under the Coupled Model Intercomparison Project Phase 6 (CMIP6) and future runs under the Shared Socioeconomic Pathway version 3 with 7.0 $\mathrm{W\cdot m^{-2}}$ radiative forcing by 2100 (SSP370). Here, the CESM-LENS2 members 91-100 [i.e., the ``mother of all runs (MOAR)''; \cite{rodgers2021ubiquity}] are selected; they are available at 6-hourly and 1.2$^\circ$-by-0.9$^\circ$ resolution; they aim to test the AI-based LAM for (1) future-climate boundary forcing generalization, and (2) reanalysis-to-ensemble boundary forcing generalization. All boundary forcing datasets are linearly interpolated from their original grid spacing to the 4-km grid. Their z-score and residual normalization coefficients are inherited from the 1980--2018 ERA5 values used in model training.

\section{Methods}\label{sec3}

\subsection{Model design and training}\label{sec31}

The AI-based LAM is a Swin-Transformer-based model with tensor embedding, resampling, and residual blocks (Figure~\ref{fig1}c). The Swin-Transformer blocks are configured with eight attention heads, window sizes of (7, 7), and Multi-Layer Perceptron (MLP) ratios of 4. The AI-based LAM has three input branches for (1) single-time, hourly high-resolution inputs from the timestep previous to the target (e.g. CONUS404 for training); (2) boundary forcing inputs on two neighboring times (e.g. 3-hourly ERA5 for training); and (3) the encoded day-of-year and time-of-day of the inputs and target inputs. A Feature-wise Linear Modulation (FiLM) layer is applied to combine the temporal information with boundary forcing inputs. The two boundary forcing time stamps are selected as the ``latest'' time window (i.e., the one that provides the most future information; Figure~\ref{fig1}f). The post-processing model is a simplified U-Net with hidden-layer sizes increasing progressively from 64 to 512 (Figure~\ref{fig1}d). 

The AI-based LAM and the post-processing model are trained separately from scratch. The AI-based LAM training consists of single-step pre-training and multi-step fine-tuning. The single-step pre-training has 120 full epochs with 3000 batches per epoch and 32 samples per batch. The AdamW optimizer \cite{loshchilov2017decoupled} is used with weight decay of $3\times 10^{-6}$, an initial learning rate of $10^{-3}$, and the half-cosine-annealing learning rate schedule. For multi-step fine-tuning, a fixed learning rate of $3\times 10^{-7}$ is applied \cite<c.f.>{lam2023learning} over 5 epochs. The number of iterative steps increases per epoch from 2 to 6, corresponding to downscaling steps of 2 to 6 hours. Training loss is mean squared error with inverse-variance weights and per-variable-level weights. The training of the U-Net post-processing model is similar to that of the AI-based LAM pre-training, but with early stopping applied. The training is conducted using PyTorch \cite{paszke2019pytorch} and the Community Research Earth Digital Intelligence Twin \cite<CREDIT; >{schreck2025community} platform on 32 NVIDIA A100 GPUs and on the Derecho system \cite{derecho} at the NSF National Center for Atmospheric Research (NCAR). 
% Further technical details of the hyperparameter search and model training are summarized in the Supporting Information.

\subsection{Experiment design}\label{sec32}

This study comprises five dynamical downscaling experiments:

\begin{enumerate}
    \item \textit{LAM-3H-ERA5} covers water year (WY) 2021--2024 (from 1 October 2020 to 30 September 2024). It uses 3-hourly ERA5 as boundary forcings, the same as in the AI-based LAM training. The purpose of this experiment is to evaluate how well the model is trained and to provide a ``no generalization'' baseline to compare with other generalized dynamical downscaling experiments.

    \item \textit{LAM-6H-ERA5} covers WY 2021--2024, similar to the LAM-3H-ERA5, but using 6-hourly ERA5. The purpose of this experiment is to evaluate how well the AI-based LAM, trained on 3-hourly ERA5, can generalize to a different boundary forcing step, which is of practical importance given the common availability of 6-hourly boundary forcings.

    \item \textit{LAM-6H-GDAS} covers WY 2021--2024 and replaces the ERA5 boundary forcings with GDAS/FNL. Comparing this experiment with the two ERA5-based runs examines how well the AI-based LAM can generalize to a coarser and lower-quality boundary forcing that it was not trained on.

    \item \textit{LAM-6H-CESM-30Y} is a two-part 30-year experiment using boundary forcings from a single CESM-LENS2 ensemble member, downscaling historical conditions for 1980--2010 and SSP370 conditions for 2070--2100. By examining the downscaled climate change signals, this experiment evaluates the ability of the AI-based LAM to generalize to different climate regimes.
    
    \item \textit{LAM-6H-CESM-ENS} downscales 10 CESM-LENS2 MOAR members over 2025--2027 under the SSP370 scenario. By comparing the ensemble mean and spread of the downscaled outputs with those of the original 10 CESM-LENS2 members, this experiment examines how applicable the AI-based LAM is to the use case of downscaling an ensemble.
\end{enumerate}

All experiments are initialized from the 2010--2019 CONUS404 climatology\footnote{We found that the AI-based LAM is flexible in its initialization choice, e.g., it can be initialized with constant values. Initializing from CONUS404 climatology is ideal because the model can spin up quickly, typically within 96 hours.} 
The present-day downscaling experiments (LAM-3H-ERA5, LAM-6H-ERA5, and LAM-6H-GDAS) are also post-processed (Figure~\ref{fig1}e) to produce diagnostic outputs including total precipitation. For future-climate-based experiments (LAM-6H-CESM-ENS and LAM-6H-CESM-30Y), the focus is on the generalization abilities of the AI-based LAM.

\subsection{Verification methods}\label{sec33}

The verification of this study covers two parts. (1) For present-day downscaling experiments, they are verified in WY 2021--2024 by using CONUS404 as targets. (2) For future-climate-based experiments, their long-term statistics and consistencies with their corresponding CESM-LENS2 boundary forcings are evaluated.

The metrics for verifying prognostic variables are the Root Mean Square Error (RMSE) and the energy spectrum. RMSE is averaged over the entire spatial domain. The energy spectrum is radially averaged, which quantifies the distribution of variance in downscaling outputs across spatial scales \cite{morss2009spectra}. It is computed using 2-D Fast Fourier Transforms with a Hann window. The focus of this evaluation is the amount of spatial variance that contributes to the 10-100 km mesoscale wavelength.

The metrics for post-processed total precipitation are the Equitable Threat Score (ETS), and object-oriented verification. ETS verifies categorical forecast accuracy by adjusting the standard Threat Score for hits that occur purely by chance (TP$_r$). ETS is defined using the confusion matrix components of True Positive (TP), False Positive (FP), True Negative (TN), and False Negative (FN) samples \cite{schaefer1990critical,hamill2006measuring}:

\begin{equation}
\text{ETS} = \frac{\text{TP}-\text{TP}_r}{\text{TP}+\text{FP}+\text{FN}-\text{TP}_r}\quad\quad 
\text{TP}_r = \frac{(\text{TP}+\text{FP})(\text{TP}+\text{FN})}{\text{TP}+\text{FP}+\text{TN}+\text{FN}}
\end{equation}

\noindent
In this study, ETS is computed from daily total precipitation at each grid point in the domain, over a 30-day time window. 

Object-oriented verification identifies precipitation spatial features as objects using thresholds of 2.54, 25.4, and 63.5 $\mathrm{mm\cdot day^{-1}}$. Object attributes, including object size, counts, and percentile-based central intensities, are verified against CONUS404. This verification allows for displacement and structural difference between downscaling outputs and targets, thereby avoiding the double penalty inherent in traditional grid-to-grid scores \cite{davis2006object}. We use the Method for Object-Based Diagnostic Evaluation (MODE) from the METplus platform \cite{metplus,brown2021model} for this verification. 
% Further technical details of the metrics above are summarized in the Supporting Information.

\section{Results}\label{sec4}

\subsection{Present-day downscaling performance}\label{sec41}

\subsubsection{Deterministic verification and long-term statistics}\label{sec411}

\begin{figure}
    \centering
    \includegraphics[width=\columnwidth]{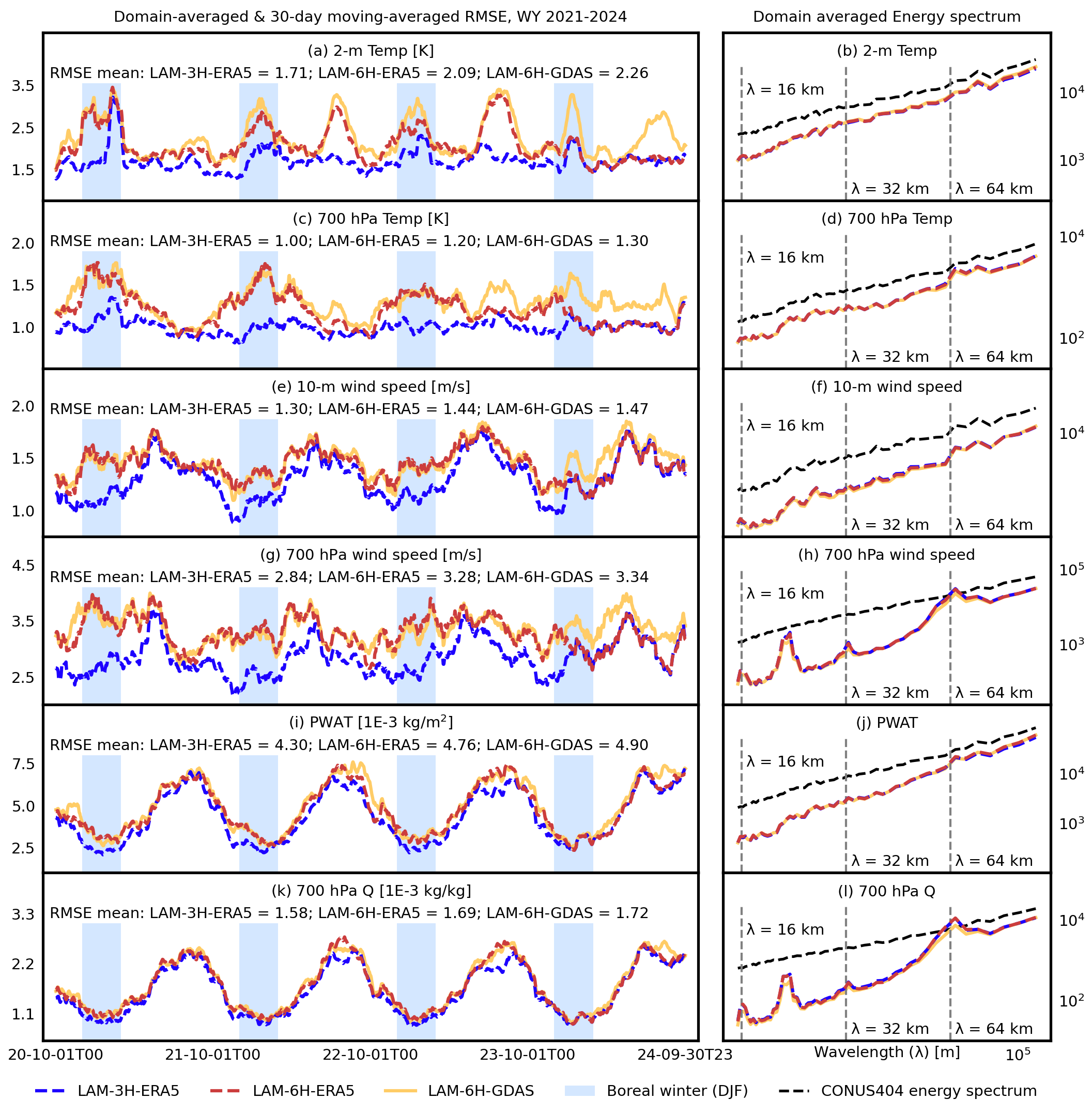}
    \caption{(a) Domain-averaged and 30-day moving averaged hourly 2-m air temperature (``2-m Temp'') RMSE in WY 2021--2024 for LAM-3H-ERA5 (blue), LAM-6H-ERA5 (red), and LAM-6H-GDAS (orange). Light blue shading indicates boreal winter. (b) The energy spectrum of 2-m air temperature in the 10--100 km wavelength range, where the black dashed line is computed from CONUS404. (c--d) As in (a--b), but for 700 hPa air temperature (``700 hPa Temp''). (e--f) As in (a--b), but for 10-m wind speed. (g--h) As in (a--b), but for 700 hPa wind speed. (i--j) As in (a--b), but for total precipitable water (``PWAT''). (k--l) As in (a--b), but for total water path (``Q'').}
    \label{fig2}
\end{figure}

All present-day downscaling experiments maintain low and temporally stable RMSE throughout WY 2021--2024 (Figure~\ref{fig2}). Over the full evaluation period of more than 35,000 hourly iteration steps, RMSE exhibits seasonal modulation but no systematic drift, indicating that the AI-based LAM can support continuous multi-year downscaling without progressive degradation in skill.

% Period-mean RMSEs are 1.71/2.09/2.26 K for 2-m air temperature and 1.17/1.45/1.58 K for 700 hPa air temperature (LAM-3H-ERA5 / LAM-6H-ERA5 / LAM-6H-GDAS)
For both 2-m and 700-hPa air temperature, RMSE is typically in the 1.0--3.0 K range with seasonal peaks (Figure~\ref{fig2}a,c). This performance is comparable to other forecasting-oriented AI-based LAMs on 48--72 hour lead times \cite<c.f.>[on the southern Great Plains]{xu2025artificial}. Updating boundary forcing steps every 3 hours yields consistently lower RMSE than the two 6-hourly configurations: by approximately 18\% relative to LAM-6H-ERA5 and 24\% relative to LAM-6H-GDAS. The two 6-hourly experiments are broadly comparable, with LAM-6H-ERA5 marginally better than LAM-6H-GDAS during parts of the summer. In addition, RMSE for 700-hPa air temperature, which represents the performance of the lower free atmosphere, is smaller than that of the 2-m air temperature; this can be attributed to the additional near-surface complexity associated with land--atmosphere coupling and boundary-layer structure.

For 10-m and 700-hPa wind speed, the domain-averaged RMSEs are 1.30--1.47 $\mathrm{m\cdot s^{-1}}$ and 2.80--3.27 $\mathrm{m\cdot s^{-1}}$ (Figure~\ref{fig2}e and g), respectively, which is slightly better than the 48-hour forecast performance as reported in \citeA{xu2025artificial}. Wind speed RMSEs show weaker purely seasonal behavior, but stronger event-to-event variability compared to air temperature. LAM-3H-ERA5 produces the lowest mean RMSEs, reducing the 10-m wind RMSEs by 10\% relative to LAM-6H-ERA5 and by 12\% relative to LAM-6H-GDAS. At 700 hPa, the mean reduction is 12\%. Differences between LAM-6H-ERA5 and LAM-6H-GDAS are modest (e.g., 0.03 $\mathrm{m\cdot s^{-1}}$ at 10 m and 0.06 $\mathrm{m\cdot s^{-1}}$ at 700 hPa in the period mean), indicating that the interval of boundary forcing steps is a larger contributor to mean skill than the choice of ERA5 versus GDAS/FNL forcings in these experiments.

Moisture errors exhibit strong seasonal cycles (Figure~\ref{fig2}i and k). Total precipitable water and 700-hPa total water path both show low RMSE in winter and higher RMSE in summer, consistent with larger moisture content and enhanced convective variability during the warm season. The three experiments track each other closely over time, and the dominant feature is the seasonal modulation rather than persistent separation among experiments. LAM-3H-ERA5 remains slightly better in the mean, improving total precipitable water and 700-hPa total water path RMSE by 10\% relative to the two 6-hourly experiments.

Figure~\ref{fig2}b, d, f, h, j, and l assess domain-averaged energy spectra in the 10--100 km wavelength range. For every variable, the AI-based LAM spectra (colored lines) lie systematically below the CONUS404 reference (black dashed lines), indicating reduced mesoscale variance in the downscaled fields. The difference is generally larger toward shorter wavelengths (near and below 16--32 km wavelength) and decreases toward longer wavelengths, where the curves partially converge. Differences between LAM-6H-ERA5 and LAM-6H-GDAS are comparatively small relative to the gap to CONUS404, implying that the mesoscale energy difference is a robust feature of these downscaling experiments, regardless of boundary forcings. %Where separations are visible, LAM-3H-ERA5 tends to retain slightly more variance than the 6-hourly experiments, consistent with its lower RMSE and suggesting that more frequent boundary updating can modestly improve downscaling performances across variables and levels.

\begin{figure}
    \centering
    \includegraphics[width=\columnwidth]{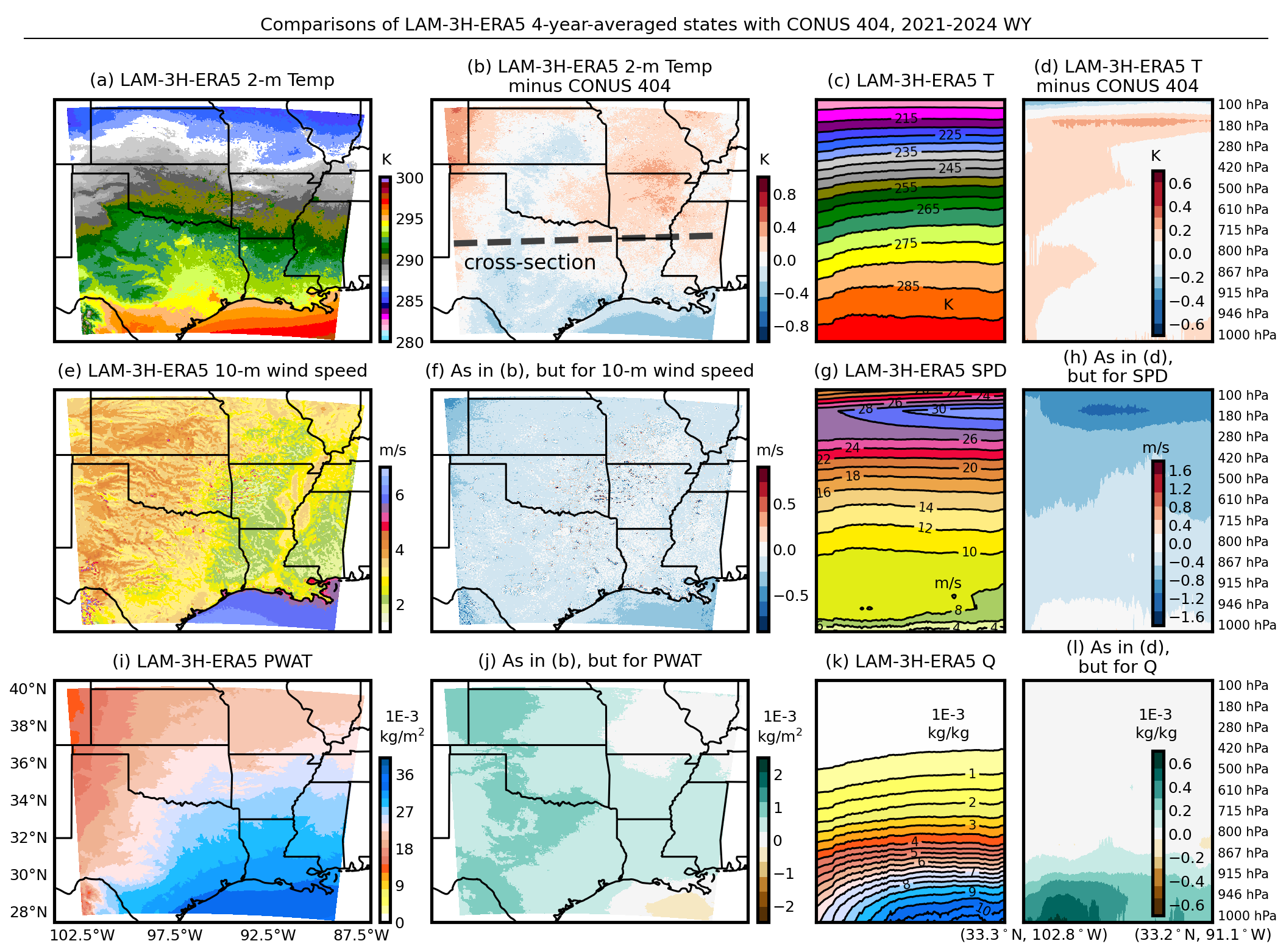}
    \caption{(a) 2-m air temperature (``2-m Temp'') produced by LAM-3H-ERA5 averaged over WY 2021--2024. (b) The difference between (a) and the corresponding CONUS404 target.(c--d) As in (a--b), but for air temperature (``T'') along the zonal cross-section from 33.3$^\circ$N, 102.8$^\circ$W to 33.2$^\circ$N, 91.1.0$^\circ$W (dashed gray line). (e--f) As in (a--b), but for 10-m wind speed. (g--h) As in (c--d), but for wind speed aloft (``SPD''). (i--j) As in (a--b), but for total precipitable water (``PWAT''). (k--l) As in (c--d), but for total water path (``Q'').}
    \label{fig3}
\end{figure}

To evaluate the fidelity of the downscaled mean state and diagnose potential climatological drift, the WY 2021--2024 from all present-day downscaling experiments are compared to the CONUS404 reference. Figure~\ref{fig3} presents the result for LAM-3H-ERA5. Overall, across air temperature, wind speed, and moisture content, the multi-year mean biases are small in amplitude and spatially coherent, clearly smaller than the corresponding hourly RMSE values in Figure~\ref{fig2}. The separation between small biases in the mean and larger instantaneous RMSEs indicates that the errors predominantly arise from transient weather variability rather than systematic drift.

The four-year mean 2-m air temperature reproduces the expected large-scale gradients and geographical signatures across the domain, including north--south and land--ocean temperature contrasts (Figure~\ref{fig3}a). The corresponding mean bias is generally less than 0.5 K in magnitude (Figure~\ref{fig3}b). A vertical cross-section of mean air temperature displays the expected tropospheric stratification, with a strong vertical gradient and comparatively weaker longitudinal variability along the transect (Figure~\ref{fig3}c). A minor warm bias is found through much of the column, generally less than 0.2 K in magnitude (Figure~\ref{fig3}d).

The four-year mean 10-m wind speed captures spatial heterogeneity well, including stronger winds over the Gulf and structured patterns over land, consistent with surface roughness and terrain influences (Figure~\ref{fig3}e). The mean 10-m wind speed bias is weak over most of the domain and is typically negative (Figure~\ref{fig3}f), indicating a slight underestimation of near-surface wind speeds relative to CONUS404. Bias magnitudes are generally well within 0.1 $\mathrm{m\cdot s^{-1}}$. The wind-speed cross-section (Figure~\ref{fig3}g) clearly captures an upper-tropospheric jet maximum at 250--100 hPa. The bias cross-section (Figure~\ref{fig3}h) shows that LAM-3H-ERA5 modestly underestimates wind speeds in the jet-core region, with negative differences of roughly 1.5 $\mathrm{m\cdot s^{-1}}$, while lower-tropospheric differences are comparatively small.

For total precipitable water, the four-year mean field shows the expected dry-to-moist gradient from the western and northern portions of the domain toward the Gulf Coast (Figure~\ref{fig3}i). The total precipitable water bias is mostly positive over much of the interior (Figure~\ref{fig3}j), with comparatively smaller and more localized dry biases near parts of the Gulf Coast. The bias amplitude remains small relative to the total precipitable water RMSE (Figure~\ref{fig2}i), again supporting the conclusion that mean-state drift is not the dominant error mode. The total water path cross-section (Figure~\ref{fig3}k) shows strong near-surface moisture and a rapid decrease with height, consistent with the vertical distribution of moisture contents. The corresponding bias (Figure~\ref{fig3}l) is concentrated mostly in the lowest troposphere, with a modest near-surface moist bias over the western portion of the transect, and near-neutral differences aloft. Overall, the cross-sections confirm that the downscaled long-term mean reproduces realistic thermodynamic and dynamical vertical structures, with only modest, vertically confined mean biases.

\begin{figure}
    \centering
    \includegraphics[width=\columnwidth]{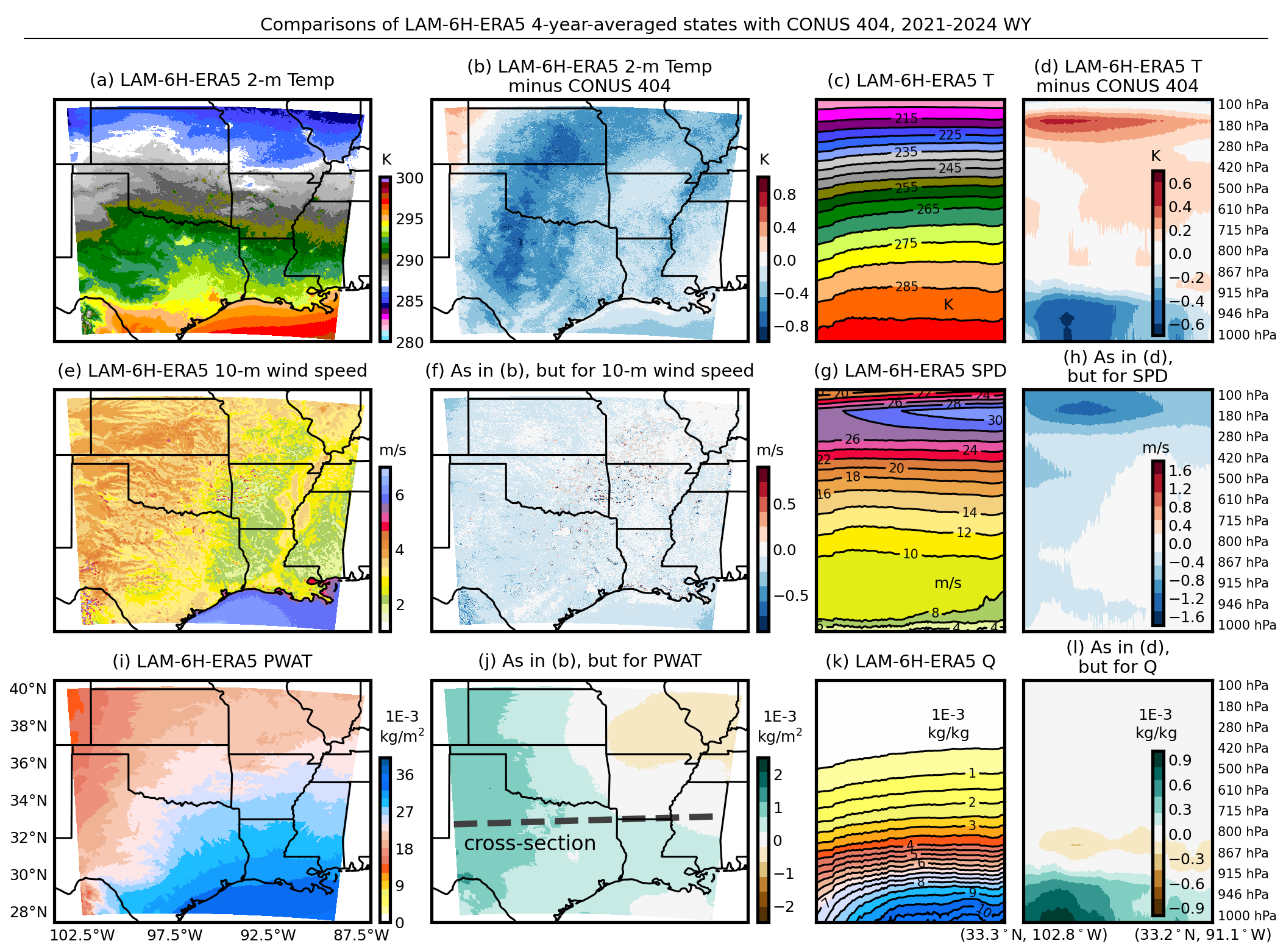}
    \caption{As in Figure~\ref{fig3}, but for the downscaling experiment of LAM-6H-ERA5.}
    \label{fig4}
\end{figure}

\begin{figure}
    \centering
    \includegraphics[width=\columnwidth]{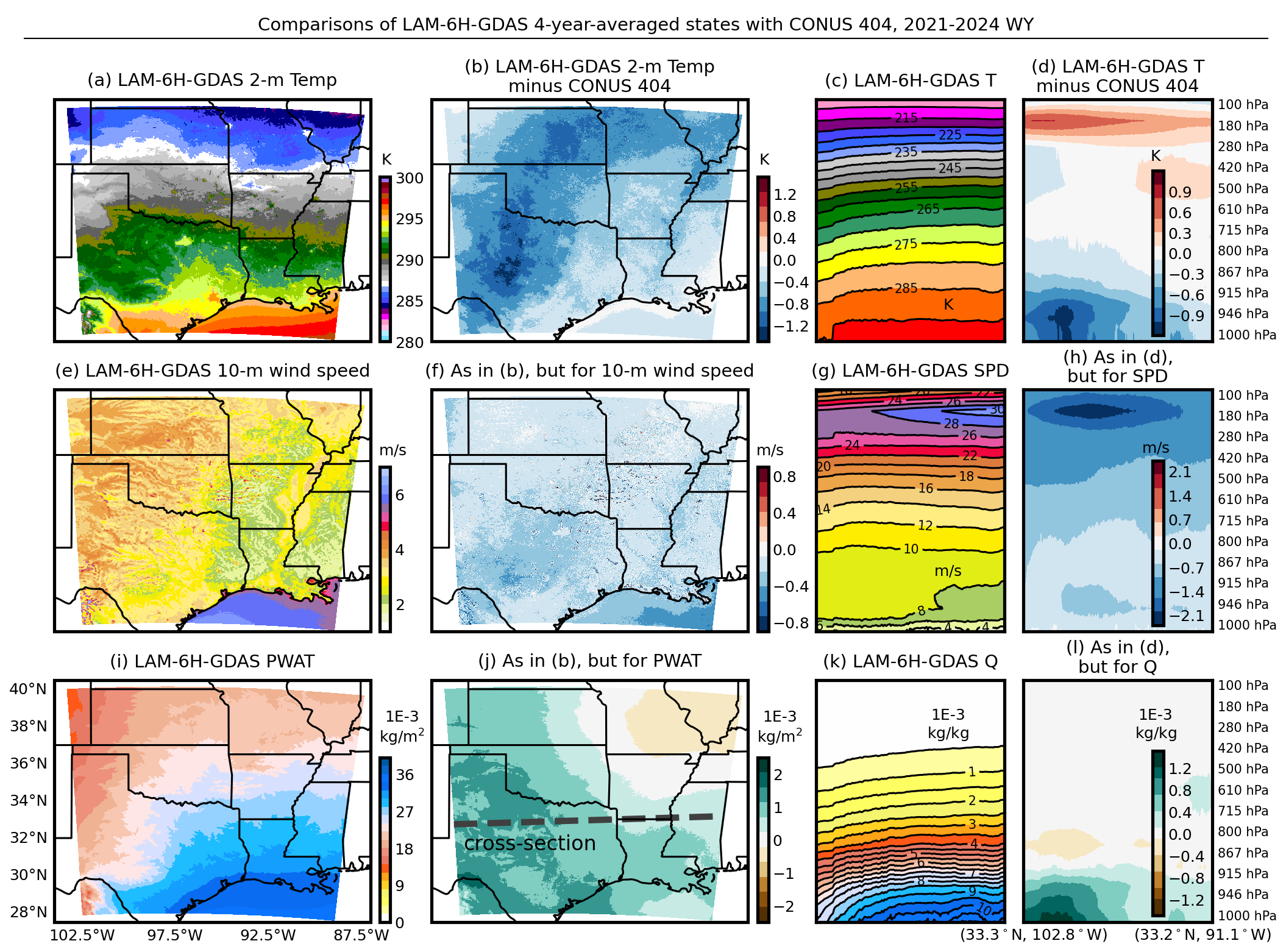}
    \caption{As in Figure~\ref{fig3}, but for the downscaling experiment of LAM-6H-GDAS.}
    \label{fig5}
\end{figure}

Figure~\ref{fig4} and Figure~\ref{fig5} show the same analysis for LAM-6H-ERA5 and LAM-6H-GDAS, respectively. In the four-year mean states, the two 6-hourly experiments show essentially the same large-scale spatial patterns as LAM-3H-ERA5 in Figure~\ref{fig3}. Differences appear in the mean bias maps, which are generally larger in magnitude. The 2-m air temperature bias shifts from a weak difference in LAM-3H-ERA5 toward a stronger interior cold bias in both 6-hourly runs. The 10-m wind speed bias remains negative in all experiments, but the underestimation is slightly stronger. For total precipitable water, LAM-3H-ERA5 shows a weak, mostly positive bias over most of the domain, whereas the 6-hourly runs exhibit larger and more structured precipitable water biases, including a dry-bias pattern over the northeastern side of the domain and a broader moist bias over parts of the southwestern domain in LAM-6H-GDAS.

The vertical structure along the transect is also qualitatively consistent across all present-day downscaling experiments in terms of the mean state. Compared with the weak, broadly warm temperature bias in LAM-3H-ERA5, both 6-hourly experiments develop a clearer warm bias near the upper troposphere around 180 hPa and a cold bias in the lowest troposphere, roughly 1000--900 hPa. In wind speeds, the jet-core underestimation in LAM-3H-ERA5 becomes broader in LAM-6H-GDAS. This joint pattern of a warm-over-cool temperature bias and weaker upper-level winds is dynamically consistent with reduced thermal-wind shear in the long-term mean. Total water path biases remain concentrated near the surface in all experiments, but the amplitude increases as the boundary forcing steps decrease, implying that the moisture mean-state differences may originate from boundary-forced moisture transport and surface/vertical mixing processes rather than from the free-tropospheric humidity structure.

\begin{figure}
    \centering
    \includegraphics[width=\columnwidth]{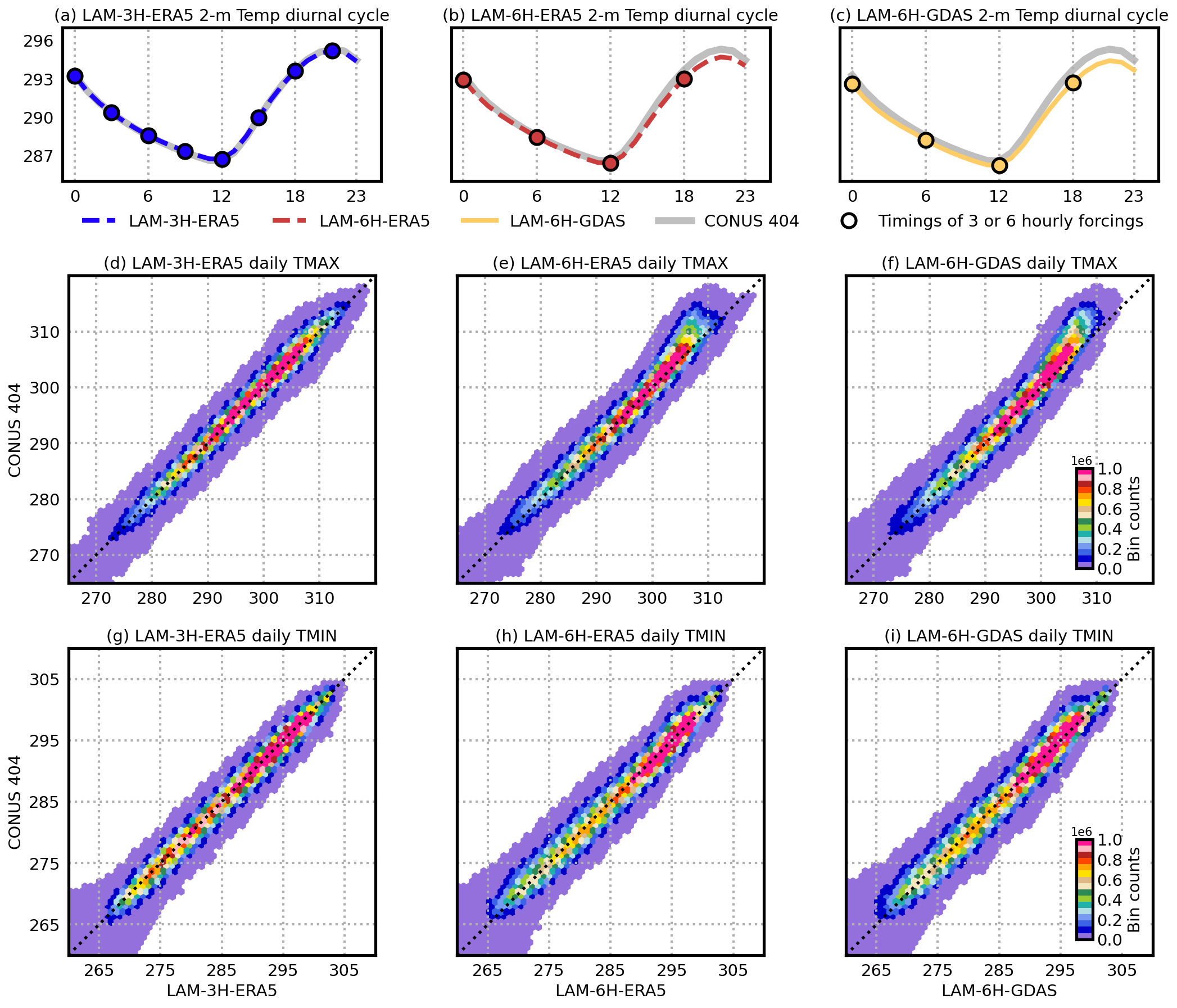}
    \caption{(a--c) WY 2021--2024 averaged diurnal cycle of 2-m air temperature (``2-m Temp'') for LAM-3H-ERA5 (blue dashed line), LAM-6H-ERA5 (red dashed line), and LAM-6H-GDAS (orange dashed line), respectively. The diurnal cycle of CONUS404 is provided as a reference (gray solid line). Circles indicate the hour-of-day that matches with 3- and 6-hourly boundary forcing steps. (d--f) 2-D histograms of daily maximum 2-m air temperature (``TMAX'') with CONUS404 on the y-axis and LAM-3H-ERA5, LAM-6H-ERA5, and LAM-6H-GDAS, respectively, on the x-axis. The identity line is shown with a black dotted line.  (g--i) As in (d--f), but for daily minimum 2-m air temperature (``TMIN'').}
    \label{fig6}
\end{figure}

Figure~\ref{fig6} further examines the diurnal cycle and distribution of 2-m air temperature to diagnose the origin of the cold bias in the 6-hourly downscaling experiments. The mean diurnal cycles (Figure~\ref{fig6}a--c) show that LAM-3H-ERA5 closely follows the CONUS404 reference throughout the day, whereas both 6-hourly experiments exhibit a systematically reduced warm-phase amplitude: the largest negative difference occurs near the daily maximum (around 21--23 UTC), while agreement remains comparatively good around the early-morning minimum (around 12 UTC). This is likely explained by the boundary forcing update timing: the 6-hourly boundary forcing steps occur at hours 0, 6, 12, and 18 UTC, which align with the diurnal minimum but do not sample the period of peak warming for the domain; by contrast, the 3-hourly boundary forcing steps include additional updates closer to the diurnal maximum. As a result, the 6-hourly experiments are less constrained by the boundary conditions during the warmest part of the diurnal cycle and tend to underestimate the diurnal temperature maximum, which translates into a cold bias in mean 2-m air temperature.

The 2-D histograms of daily 2-m air temperature maximum (Figure~\ref{fig6}d-f) indicate that all experiments perform well for moderate maxima, but the 6-hourly runs slightly under-predict the hottest days: the high temperature maxima population (top-right of Figure~\ref{fig6}e and f) shifts above the reference line, consistent with a warm-season cold bias. Because the largest temperature maxima values occur in summer, this evaluation shows that the cold bias in 6-hourly experiments is primarily a summer feature and is consistent with the summertime peaks in 2-m air temperature RMSEs (Figure~\ref{fig2}a). In contrast, the daily 2-m air temperature minimum distributions (Figure~\ref{fig6}g--i) are similar across experiments, with a tight correspondence between AI-based LAM and CONUS404.

\begin{figure}
    \centering
    \includegraphics[width=\columnwidth]{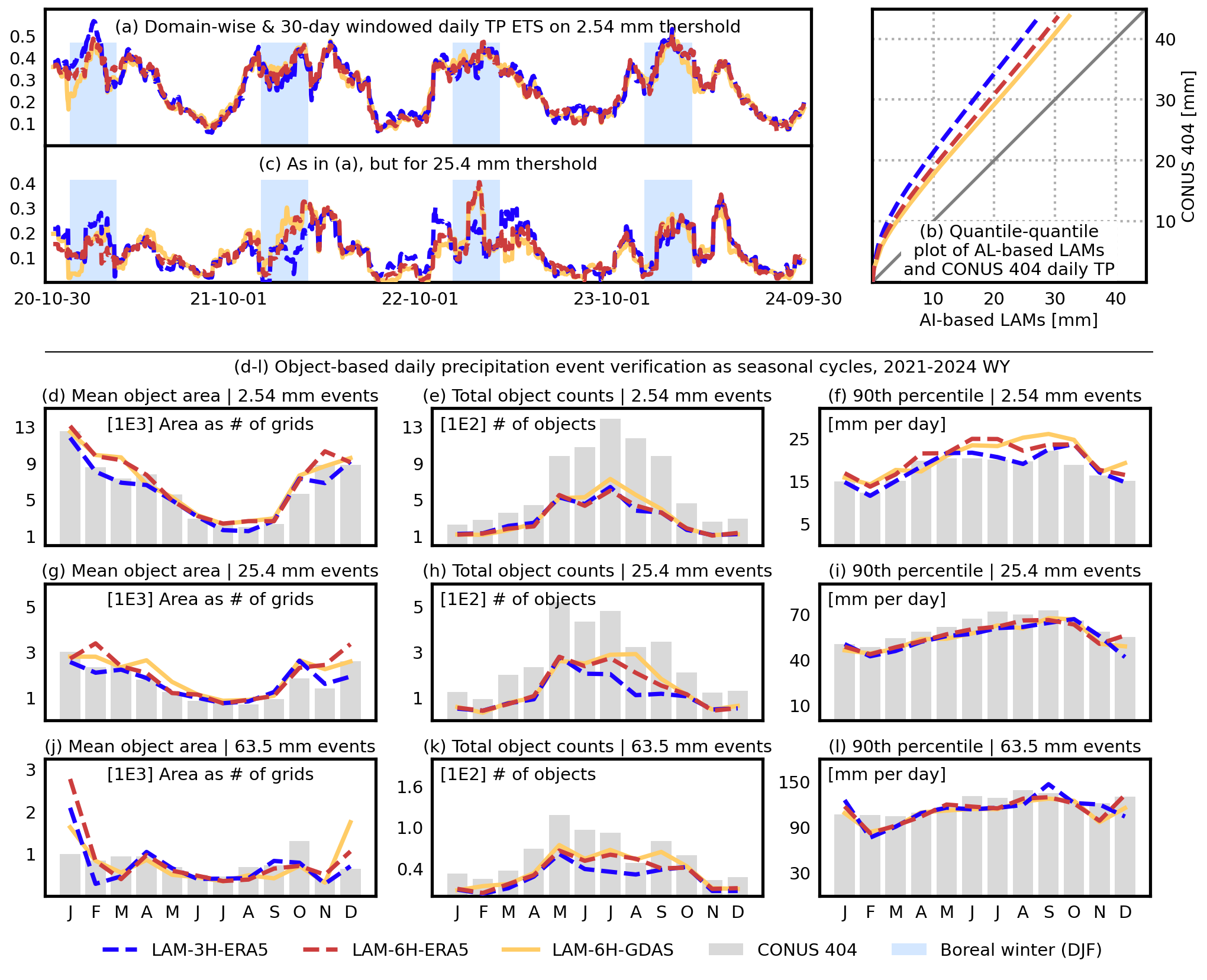}
    \caption{(a) Equitable Threat Score (ETS) for daily total precipitation (``TP'') calculated using a 2.54 $\mathrm{mm\cdot day^{-1}}$ threshold for LAM-3H-ERA5 (blue), LAM-6H-ERA5 (red), and LAM-6H-GDAS (orange). ETS is calculated over the domain and a 30-day window. Light blue areas indicate boreal winter. (b) Quantile-based comparisons of daily total precipitation between CONUS404 on the y-axis and LAM-3H-ERA5 (blue), LAM-6H-ERA5 (red), and LAM-6H-GDAS (orange) on the x-axis; gray solid line is the identity line. (c) As in (a), but using a 25.4 $\mathrm{mm\cdot day^{-1}}$ threshold. (d--f) Distributions of mean area, total number, and 90th percentile of central intensity, respectively, for precipitation objects identified using a 2.54 $\mathrm{mm\cdot day^{-1}}$ threshold; CONUS404 targets are shown as gray bars. (g--i) As in (d--f), but using a 25.4 $\mathrm{mm\cdot day^{-1}}$ threshold. (j--l) As in (d--f), but using a 63.5 $\mathrm{mm\cdot day^{-1}}$ threshold.}
    \label{fig7}
\end{figure}

Figure~\ref{fig7} verifies post-processed total precipitation from the present-day downscaling experiments against CONUS404 using categorical and object-oriented metrics. The ETS time series (Figure~\ref{fig7}a,c) indicates that all the experiments exhibit good skill in reconstructing precipitation. Overall, ETS skills are higher for the 2.54 $\mathrm{mm\cdot day^{-1}}$ threshold (Figure~\ref{fig7}a), which approximates a rain/no-rain classification, than for the 25.4 $\mathrm{mm\cdot day^{-1}}$ threshold (Figure~\ref{fig7}c), reflecting the greater difficulty of post-processing heavier daily totals. For both thresholds, ETS is systematically higher in winter, when precipitation is dominated by larger-scale synoptic systems, and lower in summer, when precipitation is more influenced by localized, convective processes. Quantile-based comparisons further show that all experiments underestimate precipitation intensity (Figure~\ref{fig7}b), and the underestimation becomes larger at higher intensities.

The object-oriented verifications (Figure~\ref{fig7}d--l) clarify how these biases arise. Across thresholds of 2.54, 25.4, and 63.5 $\mathrm{mm\cdot day^{-1}}$, all experiments can reproduce the seasonal cycles of mean object area (i.e., larger precipitation objects in winter; Figure~\ref{fig7}d, g, and j) and 90th-percentile object central intensity (i.e., stronger object central intensities in summer; Figure~\ref{fig7}f, i, and l) reasonably well, indicating that when precipitation events are reconstructed, their characteristic size and core intensity are broadly correct. However, all experiments systematically underestimate the total number of precipitation objects (Figure~\ref{fig7}e, h, and k), with the largest underestimation occurring during the warm season. Taken together with the reduced ETS skills for the heavier threshold and the underestimation of heavy precipitation quantiles, these results suggest that the post-processed precipitation successfully captures large-scale wintertime precipitation systems, but misses a fraction of the smaller-scale convectively-driven precipitation objects, primarily during summer. This under-production of warm-season precipitation objects reduces the occurrence of high daily totals and explains both the reduced ETS at the 25.4 $\mathrm{mm\cdot day^{-1}}$ threshold and the underestimation of precipitation quantiles.

\subsubsection{Case studies}\label{sec413}

\begin{figure}
    \centering
    \includegraphics[width=\columnwidth]{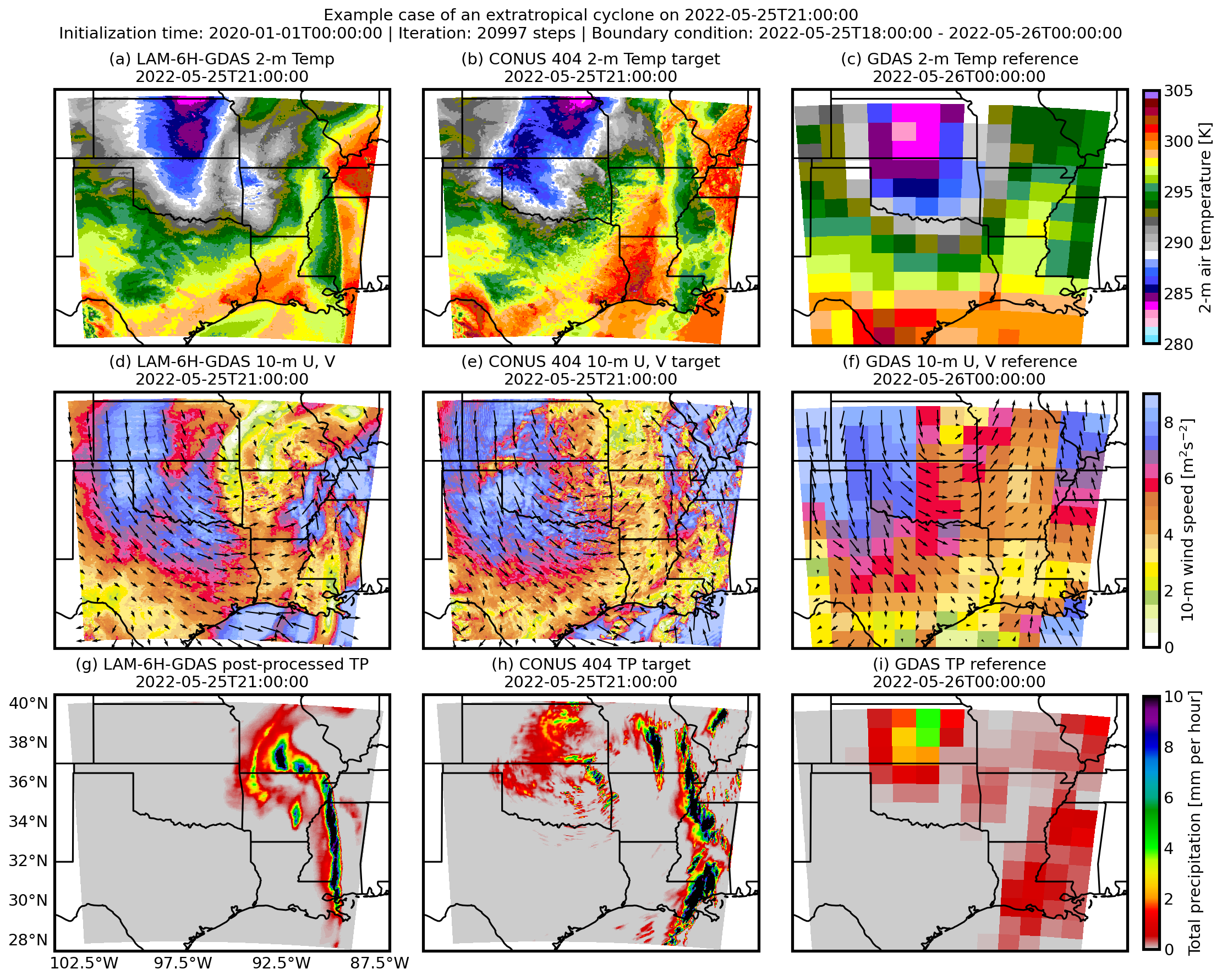}
    \caption{An example case of an extratropical cyclone from LAM-6H-GDAS on 2100 UTC, 25 May 2022, with 6-hourly boundary forcings on 1800 UTC, 25 May 2022 and 0000 UTC, 26 May 2022. (a) The downscaled 2-m air temperature (``2-m Temp'') of LAM-6H-GDAS. (b) As in (a), but for the CONUS404 target. (c) The closest 1.0$^\circ$ GDAS 2-m air temperature boundary condition. (d--f) As in (a--c), but for 10-m zonal and meridional winds (``U, V''). (g--i) As in (a--c), but for hourly total precipitation (``TP''). The GDAS total precipitation in panel (i) is rescaled from 6-hourly values and not part of the boundary forcings; it is provided as a reference only.}
    \label{fig8}
\end{figure}

% ------------------------------------------------------------------- Y.S.

Two example cases illustrate how the AI-based LAM reconstructs weather-scale systems from coarse 6-hourly boundary conditions. These case studies use LAM-6H-GDAS to represent performance with both boundary forcing step and forcing data generalizations. The first example is an extratropical cyclone characterized by a strong, cold air mass advancing from the northwest, shown at 2100 UTC on 25 May 2022 (Figure~\ref{fig8}). LAM-6H-GDAS reproduces the key thermodynamic and dynamical signatures of the frontal–cyclone system found in the CONUS404 target. In particular, the AI-based LAM resolves a sharp near-surface baroclinic zone and a comma-shaped cold-sector pattern in 2-m air temperature (Figure~\ref{fig8}a), indicating the location of occlusion and the wrapping of cold air around the cyclone center. The 10-m wind field shows a well-defined cyclonic circulation with strong winds in the advancing cold-sector flow (Figure~\ref{fig8}d), consistent with cold-air advection behind the front, and a coherent convergence signature along and ahead of the frontal boundary on the warm side (Figure~\ref{fig8}d  and e). For precipitation, the post-processing captures the placement and orientation of the primary frontal rainband (Figure~\ref{fig8}g), but may under-represent some convectively-driven precipitation in the occluded area. Overall, the AI-based LAM clearly adds high-resolution details to the coarse 1.0$^\circ$ GDAS/FNL reference, which is spatially diffuse and lacks frontal organization at the relevant scales (cf. Figure~\ref{fig8}g and i).

\begin{figure}
    \centering
    \includegraphics[width=\columnwidth]{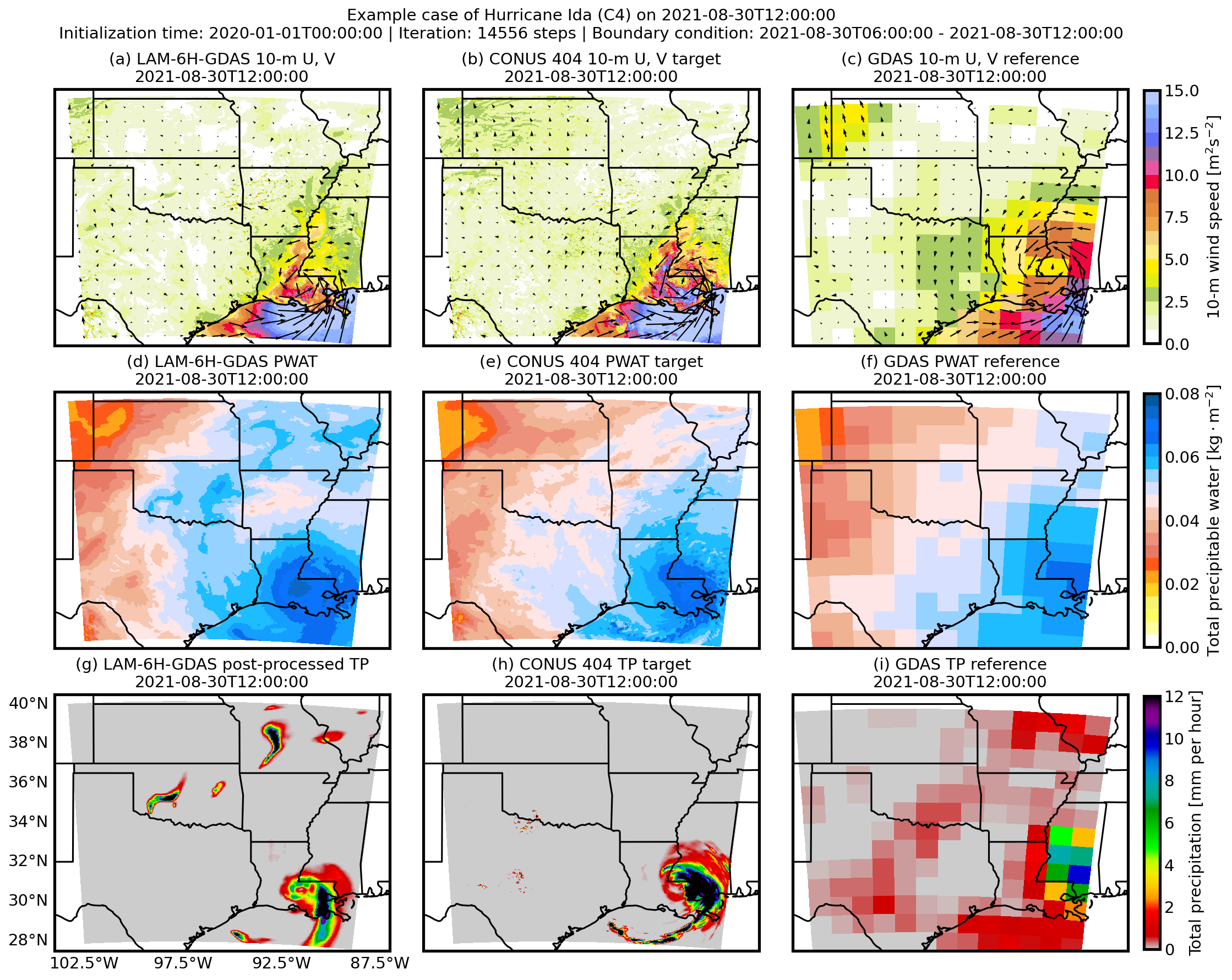}
    \caption{An example case of category 4 hurricane Ida from LAM-6H-GDAS on 1200 UTC, 30 August 2021, with 6-hourly boundary forcings at 0600 and 1200 UTC, 30 August 2021. (a) The downscaled 10-m zonal and meridional winds (``U, V'') of LAM-6H-GDAS. (b) As in (a), but for the CONUS404 target. (c) The closest 1.0$^\circ$ GDAS boundary conditions of 10-m winds. (d--f) As in (a--c), but for total precipitable water (``PWAT''). (g--i) As in (a--c), but for hourly total precipitation (``TP''). The GDAS total precipitation in panel (i) is rescaled from 6-hourly values. The GDAS total precipitable water (f) and total precipitation (i) are not part of the boundary forcings; they are provided as a reference only.}
    \label{fig9}
\end{figure}

The second example is the landing of category 4 Hurricane Ida on 1200 UTC, 30 August 2021 (Figure~\ref{fig9}). The downscaled 10-m winds reproduce the hurricane-scale circulation and the spatial structure of the near-core wind maximum relative to CONUS404 (c.f. Figure~\ref{fig9}a and b). The total precipitable water also captures the broad moist envelope surrounding the storm and the moisture gradient across the domain (Figure~\ref{fig9}d and e). A key difference between the AI-based LAM and the CONUS404 target is an overly rapid weakening after landfall. Compared to CONUS404, LAM-6H-GDAS exhibits reduced wind speeds over the land portion of the storm, which points to an overestimation of post-landfall decay. This dynamical underestimation also translates into reduced post-processed precipitation intensity and coverage over land (c.f. Figure~\ref{fig9}g and h), due to the weaker low-level moisture inflow and convergence that support rainfall production inland. Even with this difference, the post-processed precipitation retains a realistic tropical-cyclone rainband structure, and therefore clearly adds high-resolution details to the coarse 1.0$^\circ$ GDAS/FNL precipitation reference (c.f. Figure~\ref{fig9}g and i). Taken together, these examples are consistent with precipitation verifications: the framework reliably reproduces organized, large-scale precipitation structures but may miss and/or underestimate smaller-scale convective elements.

\subsection{Downscaling future climate conditions}\label{sec42}
\subsubsection{Evaluations of downscaled future climate signals}\label{sec421}

\begin{figure}
    \centering
    \includegraphics[width=\columnwidth]{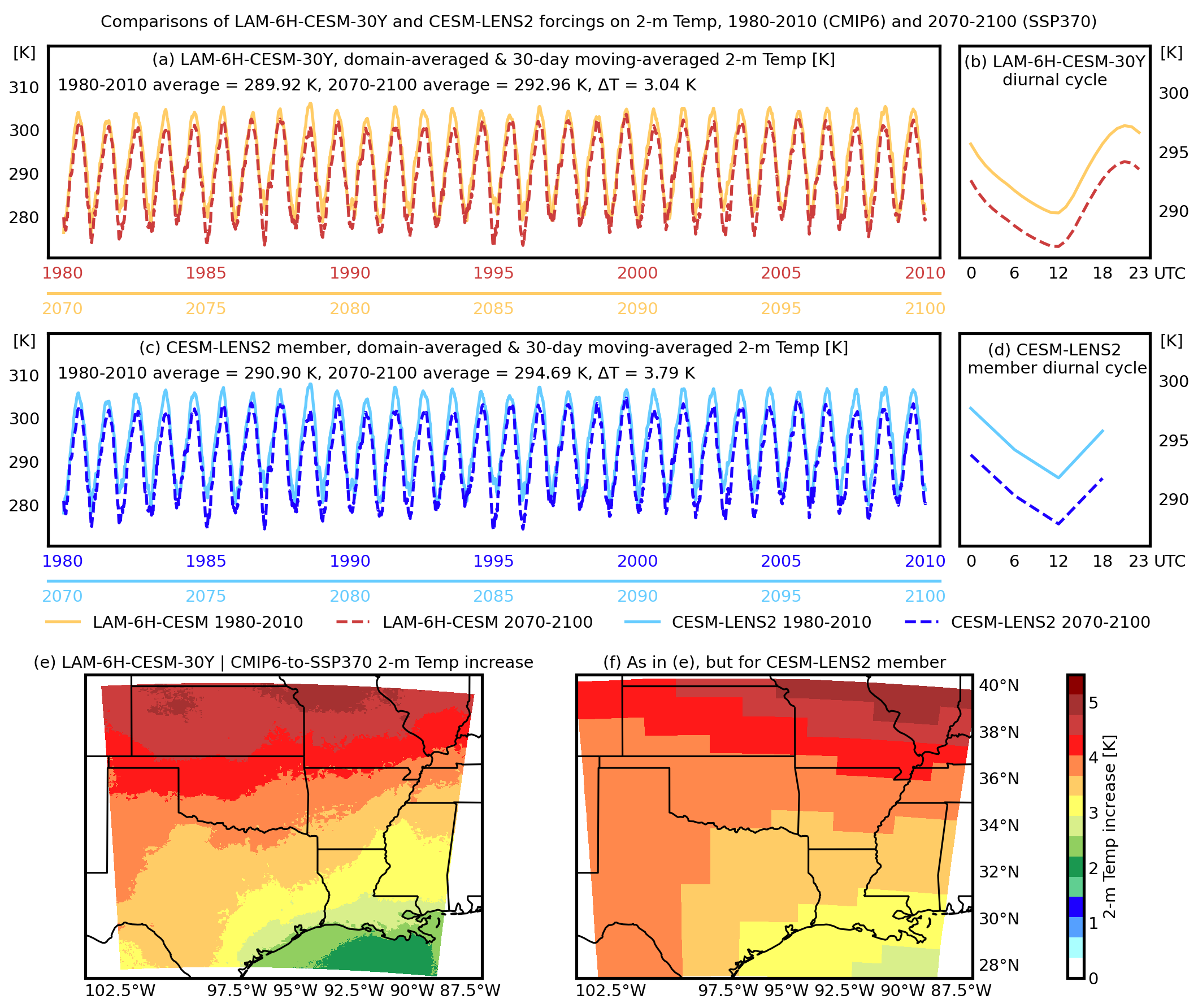}
    \caption{(a) LAM-6H-CESM-30Y 2-m air temperature (``2-m Temp''; averaged over the domain and a 30-day moving window) for the periods 1980--2010 (red dashed line) and 2070--2100 (orange solid line). (b) The diurnal cycle of LAM-6H-CESM-30Y domain-averaged 2-m air temperature. (c--d) As in (a--b), but for CESM-LENS2 2-m air temperature boundary forcings. (e) The spatial distribution of 2-m air temperature increase from 1980--2010 to 2070--2100 in LAM-6H-CESM-30Y. (f) As in (e), but for the 2-m air temperature increase in CESM-LENS2 boundary forcings.}
    \label{fig10}
\end{figure}

\begin{figure}
    \centering
    \includegraphics[width=\columnwidth]{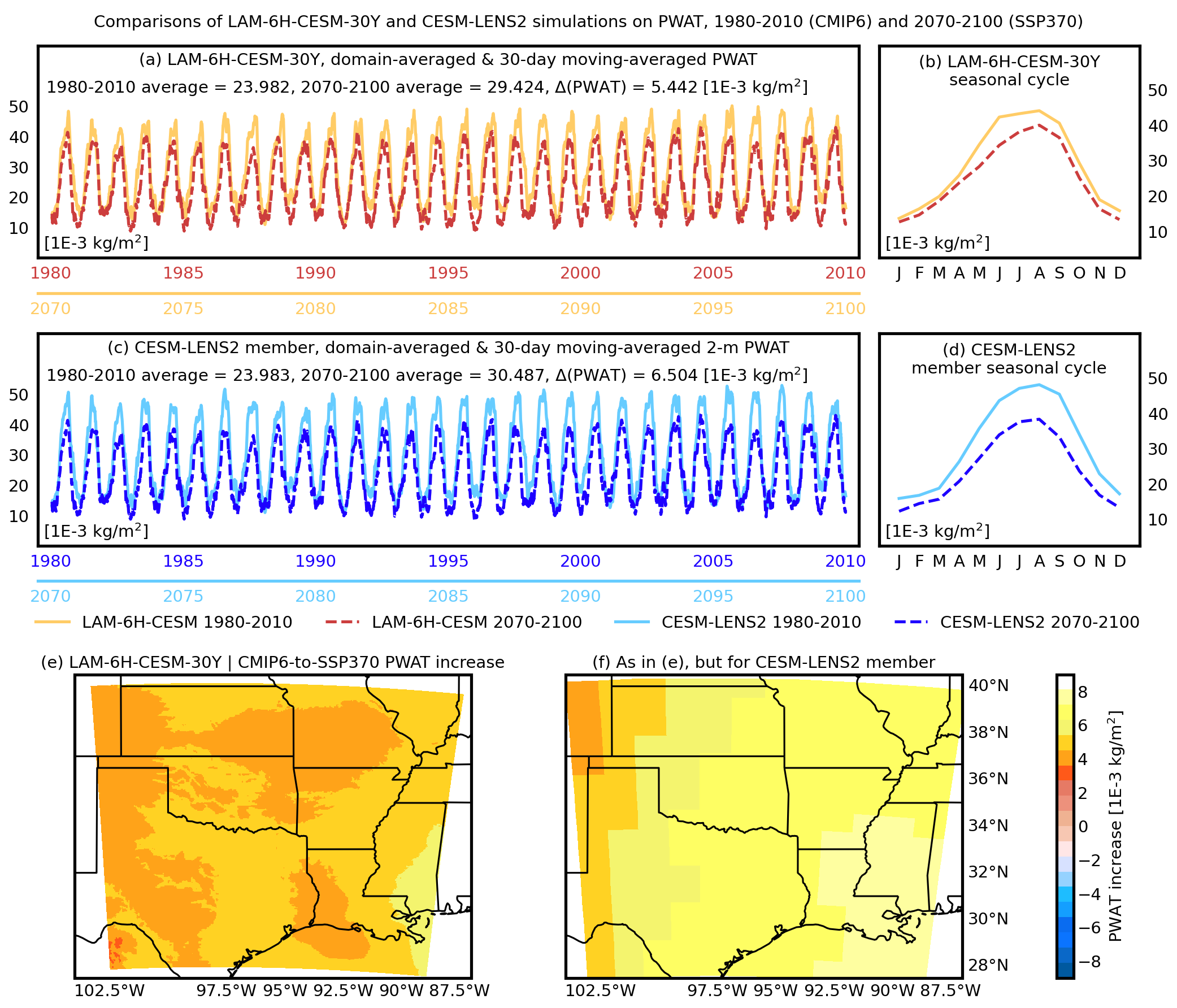}
    \caption{(a) LAM-6H-CESM-30Y total precipitable water (``PWAT''; averaged over the domain and a 30-day moving window) for the periods 1980--2010 (red dashed line) and 2070--2100 (orange solid line). (b) The seasonal cycle of LAM-6H-CESM-30Y domain-averaged 2-m air temperature. (c--d) As in (a--b), but for CESM-LENS2 total precipitable water boundary forcings. (e) The spatial distribution of total precipitable water difference between 1980--2010 and 2070--2100 in LAM-6H-CESM-30Y. (f) As in (e), but for the 30-year averaged total precipitable water difference in CESM-LENS2 boundary forcings.}
    \label{fig11}
\end{figure}

Figure~\ref{fig10} and Figure~\ref{fig11} evaluate the LAM-6H-CESM-30Y downscaling experiment by comparing 30-year historical (1980--2010 under CMIP6) and late-century (2070--2100 under SSP370) periods with boundary forcings from the corresponding CESM-LENS2 member. For 2-m air temperature (Figure~\ref{fig10}), the AI-based LAM remains stable and produces well-behaved multi-decadal downscaled outputs, even though it has only been trained using present-day ERA5 boundary conditions. 2-m air temperature, averaged over the domain and a 30-day moving window, exhibits a clean and repeatable annual cycle with no unrealistic value drift within each 30-year period (Figure~\ref{fig10}a), closely tracking the phase and amplitude of the CESM-LENS2 boundary-forcing evolution (Figure~\ref{fig10}c). The AI-based LAM also successfully reconstructs sub-daily variabilities: the domain-averaged diurnal cycle is seamlessly resolved on an hourly basis (Figure~\ref{fig10}b) while remaining consistent with the coarse 6-hourly forcing conditions (Figure~\ref{fig10}d), indicating that the model can infer realistic diurnal temperature structure from sparsely updated 6-hourly boundary conditions. The downscaled SSP370 warming signal is physically meaningful: the domain-averaged warming from 1980--2010 to 2070--2100 is 3.04 K in LAM-6H-CESM-30Y (Figure~\ref{fig10}a), comparable to the 3.79 K warming in the CESM-LENS2 boundary forcing (Figure~\ref{fig10}c). Spatially, the model reproduces the primary pattern of the warming response (Figure~\ref{fig10}e--f), including amplified warming over the northern portion of the domain of around 5.0 K. A small difference can be found over the Gulf, where the AI-based LAM indicates a 2.0 K weaker warming than the 3.0 K in the CESM-LENS2 member, which likely contributes to the smaller domain-mean 2-m air temperature difference in the downscaling outputs.

The AI-based LAM also produces stable 30-year downscaling for total precipitable water (Figure~\ref{fig11}); its domain- and 30-day-moving-averaged values exhibit a regular seasonal cycle in both periods without apparent drift (Figure~\ref{fig11}a), and capture the overall moistening from CMIP6 to SSP370 scenarios. This is consistent with the CESM-LENS2 total precipitable water increases in the late-century period. Note that, unlike 2-m air temperature, total precipitable water is not directly provided to the AI-based LAM as a boundary forcing variable (see Table~\ref{tab1}). The overall consistency between CESM-LENS2 and its downscaled total precipitable water suggests that the AI-based LAM interprets cross-variable relationships correctly. By taking future climate air temperature, humidity, and wind as boundary forcings, the model can derive total precipitable water under the same warming conditions.

The downscaled seasonal cycle of LAM-6H-CESM-30Y shows a modest underestimation of summertime total precipitable water relative to the CESM-LENS2 member (c.f. Figure~\ref{fig11}b and d). This behavior is consistent with the spatial pattern of the total precipitable water climate change signal: while total precipitable water increases broadly across the domain in both downscaling outputs and the CESM-LENS2 member, the AI-based LAM underestimates the total precipitable water increase on the eastern side of the domain (c.f. Figure~\ref{fig11}e and f). For domain-wise averages, the total precipitable water increase is $\mathrm{5.4\times10^{-3}\ kg\cdot m^{-2}}$ in LAM-6H-CESM-30Y compared to $\mathrm{6.5\times10^{-3}\ kg\cdot m^{-2}}$ in the CESM-LENS2 member (Figure~\ref{fig11}a and c), indicating that the AI-based LAM captures the sign and structure of the SSP370 moistening signal but with a modest dry bias. Overall, Figure~\ref{fig10} and Figure~\ref{fig11} demonstrate that the ERA5-trained AI-based LAM can generalize to CESM2-driven SSP370 boundary conditions, producing a stable and interpretable 30-year downscaling of both near-surface temperature and column-integrated moisture.

% --------------------------------------------------------------------------------------------- 

\subsubsection{Downscaling ensembles of future climate model runs}\label{sec422}

\begin{figure}
    \centering
    \includegraphics[width=\columnwidth]{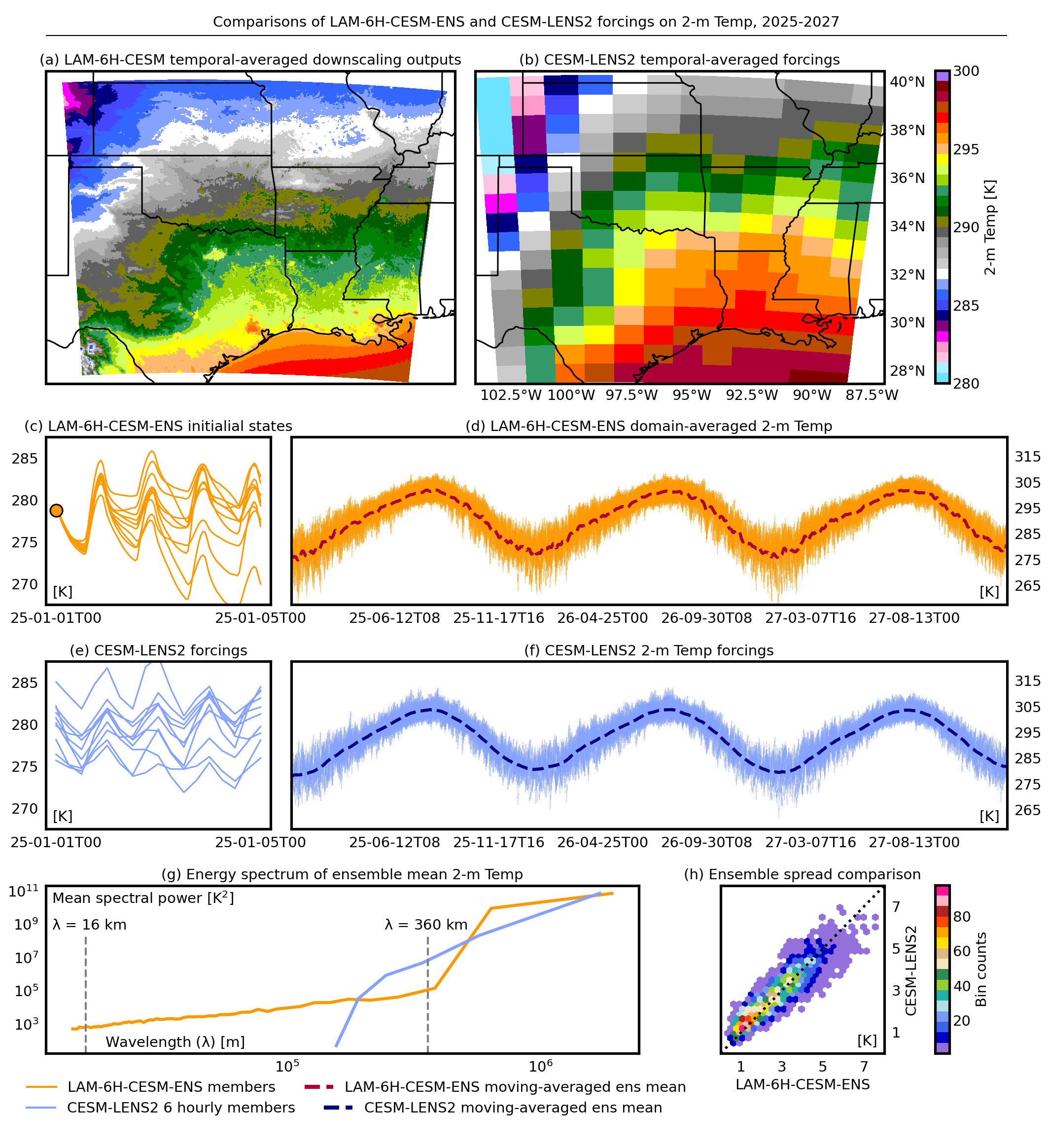}
    \caption{Evaluations of 2-m air temperature (``2-m Temp''). (a) LAM-6H-CESM-ENS downscaled and temporally averaged ensemble mean for 2025--2027. (b) As in (a), but for the CESM-LEN2 ensemble mean. (c) LAM-6H-CESM-ENS dynamically downscaled, domain-averaged ensemble members over the first 96 hours (orange solid line), with the circle representing the shared initialization of all members. (d) LAM-6H-CESM-ENS domain-averaged ensemble member (orange solid line), and ensemble mean (red dashed line). (e) The CESM-LENS2 boundary forcing ensembles that correspond to (c). (f) As in (d), but for CESM-LENS2 boundary forcing ensembles (blue solid and dashed lines). (g) The energy spectrum of CESM-LENS2 (blue solid line) and LAM-6H-CESM-ENS (orange solid line) ensemble mean. (h) 2-D histogram-based comparisons of ensemble spread between CESM-LENS2 (y-axis) and LAM-6H-CESM-ENS (x-axis). For (h), LAM-6H-CESM-ENS is downsampled from hourly to 6-hourly.}
    \label{fig12}
\end{figure}

In the ensemble future-climate downscaling experiment, LAM-6H-CESM-ENS, the AI-based LAM dynamically downscales 10 CESM-LENS2 MOAR members in 2025--2027 under the SSP370 scenario. Figure~\ref{fig12} shows that the AI-based LAM preserves the large-scale spatial distribution of the CESM-LENS2 ensemble-mean 2-m air temperature. Even after ensemble averaging, the model still adds high-resolution structure that is absent from the coarse CESM-LENS2 boundary forcings (c.f. Figure~\ref{fig12}a and b). The downscaled ensemble mean retains the same broad land-ocean and meridional gradients imposed by the CESM2 forcing, but provides detailed spatial heterogeneity associated with terrain and mesoscale variabilities, demonstrating that the AI-based LAM can downscale to fine-grained spatial information without distorting the ensemble-mean thermodynamic state.

The temporal statistics of domain-averaged 2-m air temperature are consistent with the CESM-LENS2 ensemble forcings. The downscaled members rapidly separate from a shared initialization and develop different trajectories within the first 96 hours of simulation (Figure~\ref{fig12}c), indicating that the AI-based LAM can quickly adapt to member-specific evolutions controlled by boundary forcings. Over the full 2025--2027 period, the downscaled domain-averaged time series reproduces both the ensemble mean seasonal evolution and the spread seen in the CESM-LENS2 forcings (c.f. Figure~\ref{fig12}d and f). The 2-D histogram comparing the ensemble spreads of CESM-LENS2 and LAM-6H-CESM-ENS at 6-hourly intervals clusters tightly along the identity line, implying that the AI-based LAM successfully preserves the magnitude and variability of the ensemble uncertainties (Figure~\ref{fig12}h).

Beyond mean and spread, Figure~\ref{fig12}g shows that the AI-based LAM improves the scale-dependent structure of the ensemble-mean 2-m air temperature field. The downscaled ensemble-mean energy spectrum contains more power at mesoscale wavelengths than the coarse CESM-LENS2 counterpart and exhibits generally correct inertial-range behavior with slopes consistent with the $k^{-3}$ and $k^{-5/3}$ regimes over appropriate wavelength bands. This indicates that the AI-based LAM correctly redistributes ensemble-mean spatial variability into the mesoscale, producing a more physically realistic energy spectrum than the raw CESM-LENS2.

\section{Discussion and conclusions}\label{sec5}

This study proposes an AI-based Limited Area Model (LAM) for dynamical downscaling over the Southern Great Plains and the southeastern United States. The AI-based LAM is trained using 4-km hourly CONUS404 data as the high-resolution target and 0.25$^\circ$ 3-hourly ERA5 as boundary forcings, but it is designed to be flexible with regard to boundary forcing datasets, grid spacings, and underlying climate regimes. While the AI-based LAM is focused on prognostic variables, a separate post-processing model is proposed to derive more variables, including total precipitation. The AI-based LAM and its downscaling framework are tested under various use cases, including non-ERA5 boundary conditions of 1.0$^\circ$.0, 6-hourly GDAS/FNL dataset under present-day conditions, and 0.9$^\circ$-by-1.2$^\circ$ 6-hourly CESM-LENS2 datasets under both the CMIP6 and the SSP370 scenario.

Evaluation of the downscaling experiments shows that the frequency of boundary forcing steps is a key factor in performance. Comparisons of 3-hourly ERA5 and 6-hourly ERA5 downscaling show that more frequent boundary forcing updates produce a robust performance improvement, with the most visible benefits for fields with strong diurnal-cycle components. This indicates that, if the forcing dataset provides it, a 3-hourly boundary-update strategy is preferable for regional climate applications targeting near-surface variabilities and extremes.

The AI-based LAM is found to generalize well across boundary forcing datasets and resolutions. The comparison between 6-hourly ERA5 and 6-hourly GDAS/FNL downscaling shows that, although GDAS/FNL is coarser than ERA5 and does not participate in model training, the AI-based LAM still produces high-quality downscaling outputs with comparable error amount and characteristics. Case studies further illustrate that the AI-based LAM can recover physically consistent thermodynamic gradients and circulation features (e.g., frontal baroclinicity and cyclonic flow) from 1.0$^\circ$ forcings, and that the precipitation post-processing adds high-resolution detail to the coarse GDAS/FNL precipitation fields by reconstructing organized rainband structure. These results suggest that the framework is not tightly coupled to a single reanalysis product and can be deployed with multiple operational or retrospective forcing datasets, which is important for both historical reconstruction and future planning.

Beyond present-day conditions, the downscaling experiments demonstrate that the ERA5-trained AI-based LAM can generalize to future-climate boundary conditions from CESM-LENS2 under SSP370 scenarios. 30-year downscaling outputs for both 1980--2010 and 2070--2100 remain stable and without evidence of spurious climatological drift in domain-wise diagnostics. The AI-based LAM reproduces the sign and magnitude of the late-century climate-change signals for 2-m air temperature and total precipitable water, as well as capturing their spatial patterns at regional scales. While some minor differences are found (e.g., weaker warming/moistening in parts of the eastern/ocean-influenced domain), the overall agreement indicates that the weather-scale dynamics learned by the AI-based LAM from present-day can be transferred to different climate regimes and forcings, enabling efficient regional downscaling for future scenario analysis.

The AI-based LAM also generalizes from deterministic forcing to an ensemble setting. Downscaling multiple CESM-LENS2 members shows that the AI-based LAM produces member-specific trajectories and preserves ensemble spread. The AI-based LAM can adapt quickly to member-dependent boundary information without artificially collapsing the ensemble spread. This ability is essential for regional climate risk assessment, where decision-relevant quantities require ensembles rather than single realizations.

A limitation of the AI-based LAM is its representation of mesoscale variability. Spectral diagnostics show that, relative to CONUS404, the downscaling outputs allocate less variance to the 10--100 km wavelength band, indicating an under-representation of mesoscale structures. %This limitation especially affects variables and processes whose predictability is intrinsically lower with variabilities being strongly convective-driven (e.g., warm-season precipitation). 
The object-oriented precipitation verification confirms that, while the post-processed precipitation reproduces object size and central intensity reasonably well, it underestimates the total number of precipitation objects, especially during summer, implying missed small-scale convection-driven events. Future work can potentially focus on improving the stochastic and multi-modal aspects of downscaling. One possibility is to replace the current deterministic U-Net post-processing with generative approaches (e.g., diffusion models; \cite<e.g.>{sha2025precip}) that can represent the distribution of convective structures conditioned on the large-scale state, thereby increasing the realism of mesoscale precipitation variability while maintaining consistency with the downscaled circulation.

In summary, this study provides a practical and transferable example of how AIWP models can be adapted for dynamical downscaling in a way that is computationally efficient, scientifically interpretable, and multi-year stable. By demonstrating the ability to handle flexible boundary forcing steps and usefulness across multiple forcing datasets as well as climate regimes, the proposed downscaling framework addresses key requirements for regional climate studies. Taking the AI-based LAM here as a basis, more specialized downscaling systems can be introduced to address challenges in estimating various weather and climate impacts on regional scales.

%%%%%%%%%%%%%%%%%%%%%%%%%%%%%%%%%%%%%%%%%%%%%%%
%
% DATA SECTION and ACKNOWLEDGMENTS
%
%%%%%%%%%%%%%%%%%%%%%%%%%%%%%%%%%%%%%%%%%%%%%%%

%TC:ignore
\section*{Open Research Section}
The ERA5 reanalysis data for this study can be accessed through the NSF NCAR Geoscience Data Exchange (GDEX) \cite{ecmwf2019era5} and the Google Research, Analysis-Ready, Cloud Optimized (ARCO) ERA5 \cite{carver2023arcoera5}. CONUS404 in 1980--2022 water years are available from the NSF NCAR Digital Assets Services Hub (DASH) \cite{CONUS404}. Readers may contact Dr. Lulin Xue and Dr. Aubrey Dugger via <conus404@ucar.edu> for the extended CONUS404 in 2023--2024 water years. The 1.0$^\circ$ GDAS/FNL dataset is available from GDEX \cite{NCEP2000FNL}. CESM-LENS2 datasets can also be accessed from GDEX \cite{Danabasoglu2020CESM2LE}. The AI model simulation and verification code are archived at \url{https://github.com/yingkaisha/RAL-GWC-CONUS}.
%TC:endignore

\section*{Acknowledgments}
This material is based upon work supported by the National Science Foundation (NSF) National Center for Atmospheric Research (NCAR), which is a major facility sponsored by the U.S. National Science Foundation under Cooperative Agreement No. 1852977. 
Y. Sha and A. Newman are also supported by the Department of Defense (DoD) Environmental Security Technology Certification Program award \#W912HQ24C0083 (project NH24-8403). S. McGinnis is also supported by the Department of Energy RGCM program award DOE DE-SC0016605.
The authors would like to acknowledge high-performance computing support from Derecho and Casper \cite{Cheyenne} provided by the Computational and Information Systems Laboratory, NCAR, and sponsored by the NSF.

%%%%%%%%%%%%%%%%%%%%%%%%%%%%%%%%%%%%%%%%%%%%%%%
% REFERENCES and BIBLIOGRAPHY
\bibliographystyle{apacite} % keep
\bibliography{reference}

%Reference citation instructions and examples:
%
% Please use ONLY \cite and \citeA for reference citations.
% \cite for parenthetical references
% ...as shown in recent studies (Simpson et al., 2019)
% \citeA for in-text citations
% ...Simpson et al. (2019) have shown...
%
%
%...as shown by \citeA{jskilby}.
%...as shown by \citeA{lewin76}, \citeA{carson86}, \citeA{bartoldy02}, and \citeA{rinaldi03}.
%...has been shown \citeA{jskilbye}.
%...has been shown \citeA{lewin76,carson86,bartoldy02,rinaldi03}.
%... \cite <i.e.>[]{lewin76,carson86,bartoldy02,rinaldi03}.
%...has been shown by \cite <e.g.,>[and others]{lewin76}.
%
% apacite uses < > for prenotes and [ ] for postnotes
% DO NOT use other cite commands (e.g., \citet, \citep, \citeyear, \nocite, \citealp, etc.).
%
\end{document}